\documentclass{article}

\usepackage{amsmath}

\usepackage{mathptmx}           
\usepackage{graphicx}           
\usepackage{url}                

\usepackage[a4paper,margin=2cm]{geometry}

\usepackage{natbib}
\bibpunct{(}{)}{;}{a}{,}{,}

\usepackage{textcomp}

\usepackage{xspace}

\begin{document}

\author{ Nico Reinke$^{\small{1,2,4}}$,
 Tim Homeyer$^{\small{1,2}}$,
 Michael H{\"o}lling$^{\small{1,2}}$,
  and 
  Joachim Peinke$^{\small{1,2}}$
  \\
  \\
$^{\small{1}}$ Institute of Physics,  University of Oldenburg and
\\
$^{\small{2}}$ ForWind,  University of Oldenburg, Carl-von-Ossietzky-Str. 9-11, 26129 Oldenburg, Germany
\\
$^{\small{4}}$ Corresponding author - Email: nicoreinke@gmx.de
  }
\title{Flow Modulation by an Active Grid}
\maketitle

\begin{abstract}
A new approach is shown, which estimates active grid wake features and enables to generate specific dynamically changing flow fields in a wind tunnel by means of an active grid. 
For example, measurements from free field can be reproduced in a wind tunnel.
Here, the approach is explained and applied.
Moreover, the approach is validated with wind tunnel measurements,
in terms of time series and stochastic features. 
Thus, possibilities and limitations become obvious.
We conclude that this new active grid wake estimating approach deepens the knowledge about specific flow modulations and shows new working ranges of active grids.
\end{abstract}

\section{Introduction}
Flows are omnipresent in nature.
Flow features affect functionality of technical applications as well as today's weather.
Large scale flows (e.g. atmospheric boundary layer) exhibit commonly unsteady and turbulent flow features. 
For instance, wind energy converters are strongly affected by the atmospheric turbulence and work under these changing flow conditions over many years. Thus, strongly alternating loads occur over time, which leads to unexpected fast fatigue, cf. \cite{Tavner2013}.
\\
However, the physics behind the motion of fluids are challenging and in particular the turbulent flow motion, which is by far not fully understood. 
Due to the mathematical complexity of the flow governed set of equations (Navier-Stokes equations) theoretical and numerical works are limited.
Therefore, wind tunnel experiments are often used to get an enhanced understanding of flow dynamics and turbulence. 
\\
Hence, turbulent flows and their various features play an important role in these experiments. 
Since more and more phenomena 
and technical applications 
are investigated under turbulent conditions, the generation of turbulent flows 
has become increasingly important.
Over the last years, a rising number of research groups control their turbulent flow conditions with so-called active grids, e.g.
\cite{Makita1991,Mydlarski1996,Poorte2002,Brzek2009,Cekli2010,Kamada2011,Hearst2012,Reinke2013,Bodenschatz2014,Thormann2014}.
These grids are able to generate turbulent flows with tunable features and in particular with high turbulence intensity. 
Commonly, an active grid modulates the wind tunnel flow with its dynamically moving flaps at the inlet of the test section. 
The active grid flaps are mounted next to each other on vertical and horizontal axes, whereby these axes can be individually rotated by motors. 
Beside this \textit{standard} active grid design, different approaches with other or even more degrees of freedom exist, e.g. \cite{Bodenschatz2014,Hearst2015}.
\\
The most common active grid design is suggested by \cite{Makita1991}
and wins out over other dynamic turbulence generators, e.g. \cite{Ling1972,Teunissen1975}. 
In his pioneer work \cite{Makita1991} found out that actively or dynamically disturbed flows show much stronger turbulent features than static disturbed flows, e.g with static grids, cylinder, etc..  
\cite{Mydlarski1996} point out that the motion pattern of the active grid flaps in time is essentially responsible for turbulent features. 
At first, \cite{Poorte2002} presented a motion pattern, which leads to an approximately homogenous isotropic turbulent flow.
\\
More recent experiments with active grids continue the work on motion patterns and try to find patterns which generate specific wind tunnel flow fields.
For example, \cite{Cekli2010} presented how a logarithmic shaped velocity profile can be generated by means of an active grid. 
Another example is the work done by \cite{Knebel2011}, who generated wind fields with certain spectral features which are similar to those found in the atmospheric boundary layer.
\\
Due to the enormous phase space of motion parameter, commonly, large parameter studies are necessary until a motion pattern or a control strategy is found for the flaps, which generates a specific flow, e.g. stressed in \cite{Cekli2010}. Thus, an approach would be beneficial which enables to estimate flow features to specific motion patterns of active grid flaps.
\\
This paper presents such a new approach. 
The approach is based on a characterisation of the active grid wake in its static and dynamic features. 
The gained knowledge returns necessary information about the connection of 
flow modulation and wake features.
In particular, it will be shown that the wake and its features at one fixed downstream position can be characterised in first approximation by a {static wake calibration}.
Furthermore, the presented approach continues previous concepts, e.g. \cite{Knebel2011}, where the active grid wake is reduced to the temporal mean velocity $\langle u (\alpha)\rangle_t$ as a function of angle of attack $\alpha$ of the flaps, 
which can be seen as a transfer function between $\alpha$ and $\langle u \rangle_t$.
Here, in addition velocity fluctuations $u'$ are considered, 
thus such a transfer function describes the relation between $\alpha$ and $u=\langle u \rangle_t+u'$,
which enables to generate so-called synthetic time series. 
These synthetic time series are an estimate of the active grid wakes, which belong to a specific motion pattern. Moreover, the approach enables to find a motion pattern, which generates a specific flow, i.e. specific time series or flow feature.
\\
The paper is organized as follows: Section \ref{sec:Experimental setup} shows our experimental setup.
In \mbox{section \ref{sec:Method},} the active grid wake is investigated in its static and its dynamic features. 
\mbox{Section \ref{sec:Examples}} presents three examples, which are used to explain, to apply and to validate the wake estimating approach.
Section \ref{sec:Discussion} concludes the paper. 
The appendix, section \ref{sec:Appendix}, gives a note on our new active grid design.


\section{Experimental setup}
 \label{sec:Experimental setup}
\begin{figure}
  \centering  
\includegraphics[width=0.9\textwidth]{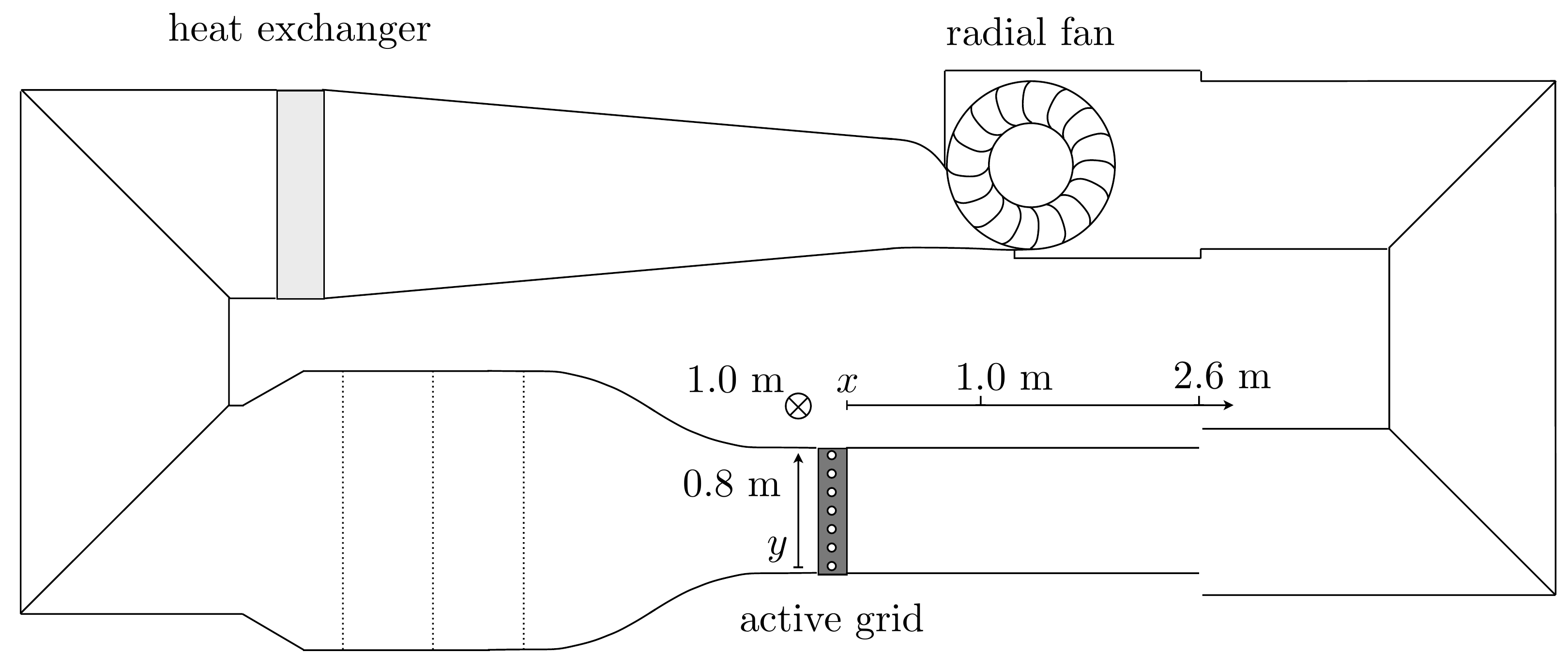}
\vspace{1cm}
 \caption{Sketch of the wind tunnel with closed test section, downstream position $x$ as well as the active grid}
\label{fig:WT}
\end{figure}

\begin{figure}
  \centering  
\includegraphics[width=0.9\textwidth]{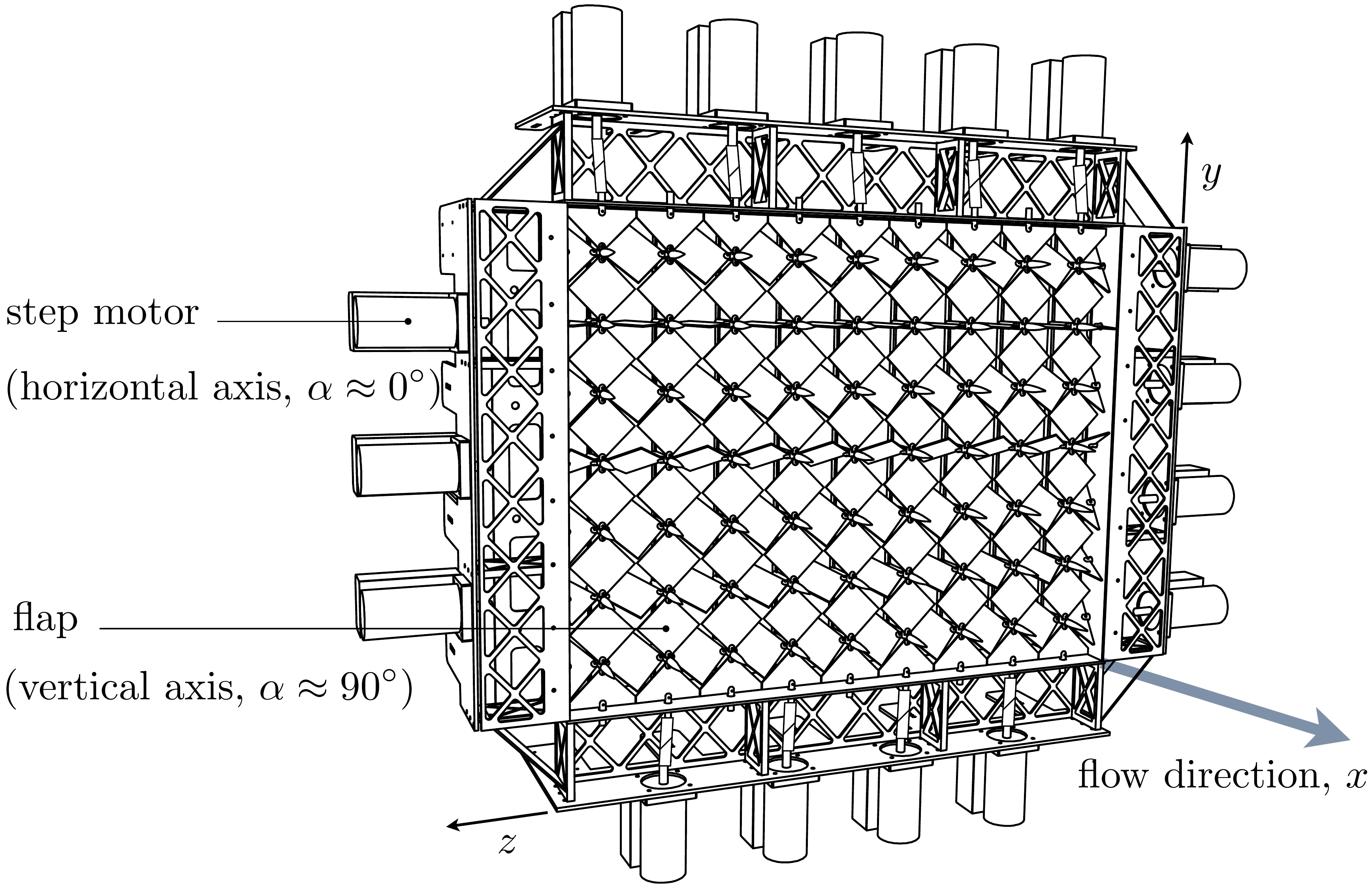}
 \vspace{0.5cm} \caption{Technical drawing of the active grid, which faces the test section}
\label{fig:aG}
\end{figure}

 Experiments are performed in a closed return wind tunnel, with test section dimensions of
$2.60\times1.00\times0.80$ m$^3$ (length $\times$ width $\times$ height), see figure \ref{fig:WT}.
Wind tunnel background turbulence intensity $I$ at the test section centerline is below 0.3\% and with mounted active grid \mbox{$I(x=1$ m$)\ge1.7\%$.} 
Inflow velocity $u_{\infty}$ is set according to table \ref{table:data}. 
Due to a heat exchanger air temperature is kept at $\tilde{T}=20^\circ$C {$\pm \ 0.5^\circ$C.} 
\begin{table}
\centering 
  \begin{tabular}{c| c| c} 
  Section                       &   $x$ [m]                    &     $ u_{\infty}$ [$\frac{\rm{m}}{\rm{s}}$]    \\
  \hline
  \ref{sec:Static}            & 1                                & 25  \\
 \ref{sec:Dynamic}        &0.1 $\leq x \leq$ 0.9   &  10  \\
   &($\Delta x = 0.1$)  & \\
\ref{sec:Examples_1}   & 0.6                             &  10  \& 25  \\
\ref{sec:Examples_2}   & 0.6                             &  10  \& 25 \\
\ref{sec:Examples_3}   & 1                             &  14   \& 25  \\
\end{tabular}
\vspace{1cm}
\caption{Text sections and corresponding measurement positions $x$ ($\Delta x$ is the distance between measurement positions)  as well as inflow adjustments $u_{\infty}$}
\label{table:data}
\end{table}
\\
Our used active grid is shown in figure \ref{fig:aG}. It consists of seven horizontal and nine vertical axes with in total 126 $0.075\times0.075$~m$^2$ square flaps. 
The mesh size $M=0.11$~m is defined as distance between two adjacent axes. 
Each of the axes can be individually held or rotated by a step motor. 
The motors allow a rotational speed up to $u_{rot}=900^\circ$s$^{-1}$ with an angular precision $\Delta \alpha \le 0.1^\circ$. The motions of the motors are determined by a user-defined motion protocol (or pattern). 
The protocol defines for the motors every $0.01$~s an angular position.  
The angular position of the flaps $\alpha$ is defined as follows, $\alpha=0^\circ$ or $180^\circ$ flaps stand parallel to the incoming wind, thus, the flaps induce a low pressure drop and a low reduction of flow speed. $\alpha=90^\circ$ or $270^\circ$ flaps stand perpendicular to the incoming wind, thus, flaps induce a high pressure drop and a high reduction of flow speed. The grid is named open when \textit{all} axes have $\alpha=0^\circ$, in this case the grid blockage is approximately $6\%$. 
The grid is named closed when \textit{all} axes have $\alpha=90^\circ$, then the grid blockage is $89\%$. 
More information about the grid can be found in the appendix, \mbox{section \ref{sec:Appendix},} where its design is compared to the design proposed by \cite{Makita1991}. 
In particular, its minimal blockage is discussed, which enhances the range of turbulence that can be controlled by the grid.
\\
Thus, during an experiment the blockage of the grid can change from $6\%$ to $89\%$ within  0.1~s, which leads to extreme loads on the grid as well as on the wind tunnel. To avoid such extreme loads a so-called {constant blockage configuration} is always used in this study. 
For the constant blockage configuration the {used} axes are split into two equal sized groups. 
The first group of axes moves according to the motion protocol and the second group rotates like the first apart from a constant phase shift of 90$^\circ$. 
Therefore, blockage changes of the first group are compensated by the second group, i.e. if the first group of axes stands open, the second group is closed. 
Consequently, the total grid blockage $\langle B \rangle_{grid}$ changes in this constant blockage configuration only weakly over time, $\frac{d\langle B \rangle_{grid}}{dt}\approx0$, compared to a non-compensating motion protocol. 
\\
For the  discussed approach below in combination with the constant blockage configuration, it is important to spatially separate the first and the second group of axes. 
Below, the inner axes (first group) modulate the flow and
the outer two side axes (second group with eight axes) compensate the motion of the inner axes according to the constant blockage configuration. 
Note that the discussion is concentrated on the first group and remarks concerning blockage etc. are related to this group.
Another characteristic of the used protocols is that the motion of axes is synchronically.
Thus, all axes which belong to one group, perform exactly the same movements and the flow gets equally modulated from one group of axes. 
\\
We would like to give to two comments on the constant blockage configuration.
First, not all axes of the grid have to be used. For instance, in some cases it is beneficial to use only vertical axes and to fix horizontal ones. 
Second, the following approach to characterise the active grid wake is not limited to the constant blockage configuration.
\\
Velocity measurements are performed by constant temperature anemometry. 
The material of hot wire is platinum-plated tungsten, its sensing length is about 2~mm and its diameter is $5~\mu$m.
A \emph{StreamLine} measurement system by \emph{Dantec} in combination with CTA Modules 90C10 and the \emph{StreamWare} version 3.50.0.9 is used for the measurements.  The overheat ratio is set to 0.8. 
Velocity measurements are performed close to the centerline in various downstream positions $x$ according to \mbox{table \ref{table:data}}.
The sampling frequency is set to $f_s=20$~kHz and the signal is low-pass filtered at $f_l=10$~kHz. The frequency response from the \emph{StreamWare} standard square wave test is approximately 27~kHz at flow speed of $u_\infty=10~\frac{\rm{m}}{\rm{s}}$.
The analog signal is converted to a digital one by \mbox{16~bit} AD-converter \emph{NI PXI 1042} \& \emph{6143}  and 
4 million samples are collected per measurement.
\\
Three calibrated single-hot-wire probes are used for reasons of mutual validation. 
The spacing between the three hot-wire probes is $\Delta y = 2.6$~cm, $y$-direction perpendicular to $x$-direction.
With these three hot-wire probes crosswise flow gradients can be characterised, which corresponds to a single flap. 
As shown in \cite{Reinke2013} 
such crosswise flow gradients decrease to a negligible level at approximately 4-5 mesh sizes downstream.

\section{Active grid wake}
\label{sec:Method}
The wake of the active grid can be separated in a {quasi static flow field}, where the wake is only a function of the flaps positions  and a {dynamic flow field}, where the flow additionally depends on the speed of flow modulations. 
This section first presents features of the static wake and secondly, an investigation concerning dynamics of flow modulations.

\subsection{Quasi static wake: velocity and fluctuations}
\label{sec:Static}
\begin{figure}
  \centering  
  \includegraphics[width=0.6\textwidth]{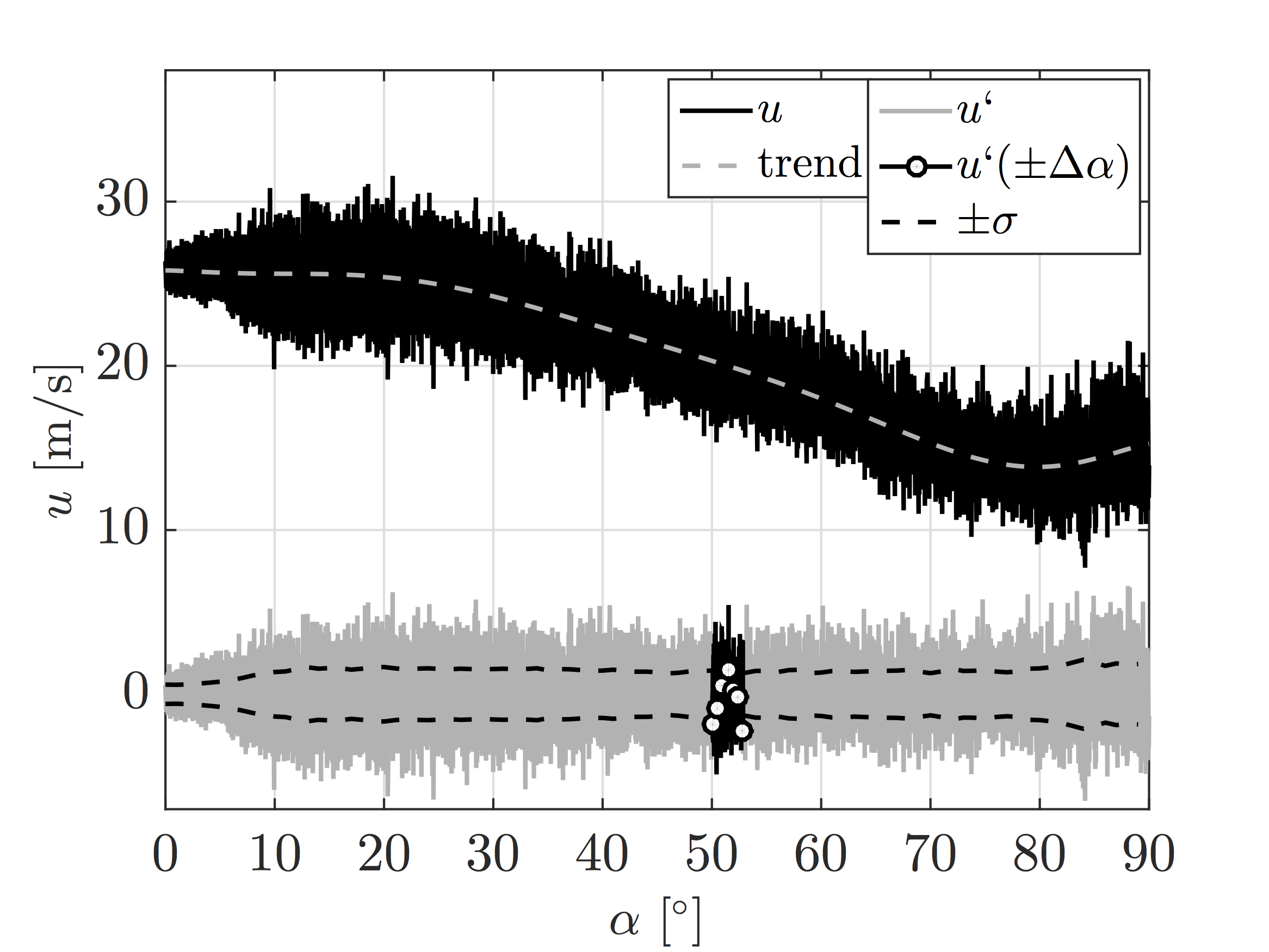}
  \vspace{1cm}
 \caption{Characteristics of the {quasi} static wake of active grid. Shown are change of velocity $u$ as a result of changing grid blockage ($\sim\alpha$), its trend, the fluctuations of velocity $u'$,  one piece of fluctuations $u'(\pm \alpha)$ and moving standard deviation of velocity $\pm \sigma(u)$}
\label{fig:AG_Chara_wake}
\end{figure}

The wake of an active grid can be stated as static when the flaps stay in a fixed position. A resulting wake is comparable with a wake of a static grid, which has a well-defined mean velocity and stationary statistic features.
Since turbulent fluctuations are still present, we call such a wake quasi static.
\\
Figure \ref{fig:AG_Chara_wake} shows how the velocity of wake $u$ changes when the grid increases its blockage and $\alpha$ of the flaps, here according to the constant blockage configuration.
Note that the velocity change in figure \ref{fig:AG_Chara_wake} is realized by very slowly changing $\alpha$ in time, from  $0^\circ$ to $90^\circ$ in 45~s. 
This velocity change is stated as static, since mean velocity ($\approx12~\frac{\rm{m}}{\rm{s}}$ - $\approx25~\frac{\rm{m}}{\rm{s}}$) is roughly four orders of magnitude higher than rotation speed of the flaps tips ($\approx0.002~\frac{\rm{m}}{\rm{s}}$).
\\
With increasing $\alpha$ velocity decreases. 
Beyond the presented range of $0^\circ \leq \alpha \leq 90^\circ$, the velocity variation continues symmetrically around $0^\circ$ and $90^\circ$.
Figure \ref{fig:AG_Chara_wake} shows a trend of the velocity change. 
The trend is determined by a polynomial fit of 8th order, which matches nicely to the change of velocity. 
Velocity fluctuation $u'$ is determined by subtracting the trend from the velocity. 
Furthermore, the windowed standard deviation of velocity $\sigma(u)$ is shown.
The number of the window samples is approximately 1000, which corresponds to step size of approximately $1^\circ$.
$\sigma$ indicates that $u'$ increases in its magnitude from $\alpha=0^\circ$ to $\approx12^\circ$; no significant change of $\sigma$ can be seen from $12^\circ$  up to $80^\circ$; between  $\alpha\approx80^\circ$ to $\approx90^\circ$ a weak increase of $\sigma$ is observed. 
\\
The trend in figure \ref{fig:AG_Chara_wake}
can be used as a transfer function,
which gives the connection between velocity, velocity fluctuations, mean velocity as well as $\alpha$.
For instance, if the velocity fluctuation at $\langle u \rangle=20~\frac{\rm{m}}{\rm{s}}$ is needed, one reads from the trend the corresponding $\alpha$, here $\alpha\approx51^\circ$ and finds simultaneously the corresponding velocity fluctuation. 
A corresponding piece of velocity fluctuations $u'(\pm\Delta \alpha)$ is marked in figure \ref{fig:AG_Chara_wake}.
$u'(\pm\Delta \alpha)$ can be also related to a certain time interval of $\Delta t$ (here $\sim1$~s) by considering sampling frequency.
Below we work with similar pieces of velocity fluctuations with a temporal length of $\Delta t = 0.5$~s. 
Note that the length of a piece can be arbitrarily chosen.
For our purpose it should be 
at least so long that the standard deviation is approximately constant within a piece. 
\\
The characterisation of the wake of the active gird, as done in figure \ref{fig:AG_Chara_wake}, is called \textit{wake calibration}.
The wake calibration gives information about possible flow features in the wake of the active grid. For instance, the minimum and maximum of the trend estimates the minimally and maximally accessible mean velocity. In case another range of velocity is required, $u_\infty$ has to be adjusted and a new wake calibration has to be performed.


\subsection{Dynamic wake: decay and interactions of flow modulations}
\label{sec:Dynamic}
\begin{figure}
  \centering  
\includegraphics[width=0.6\textwidth]{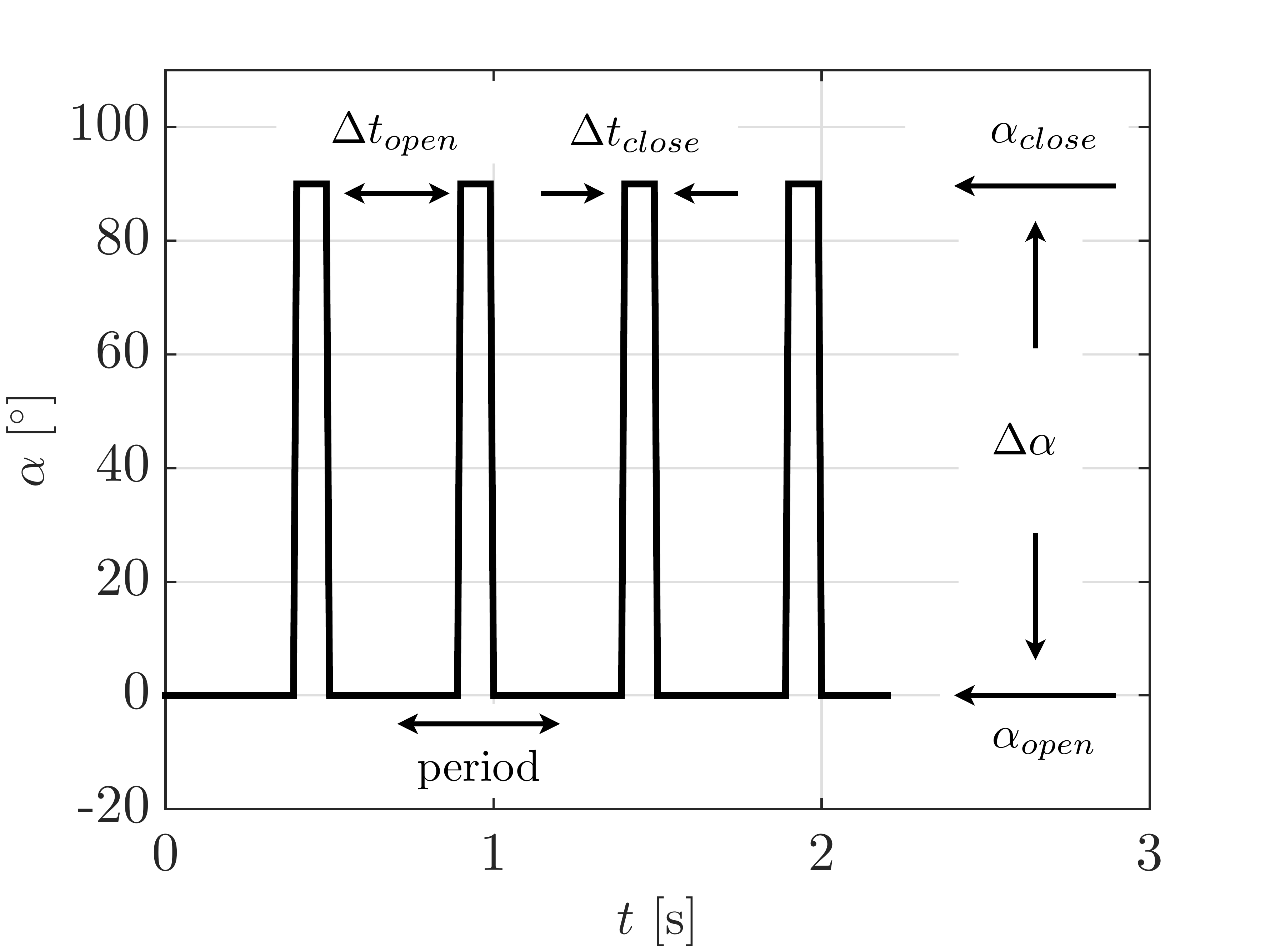}
 \vspace{0.5cm} \caption{Illustration of flap motion to generate specific flow modulations and determining modulation parameters}
\label{fig:Pulse_AoA}
\end{figure}

\begin{figure}
  \centering  
\includegraphics[width=0.6\textwidth]{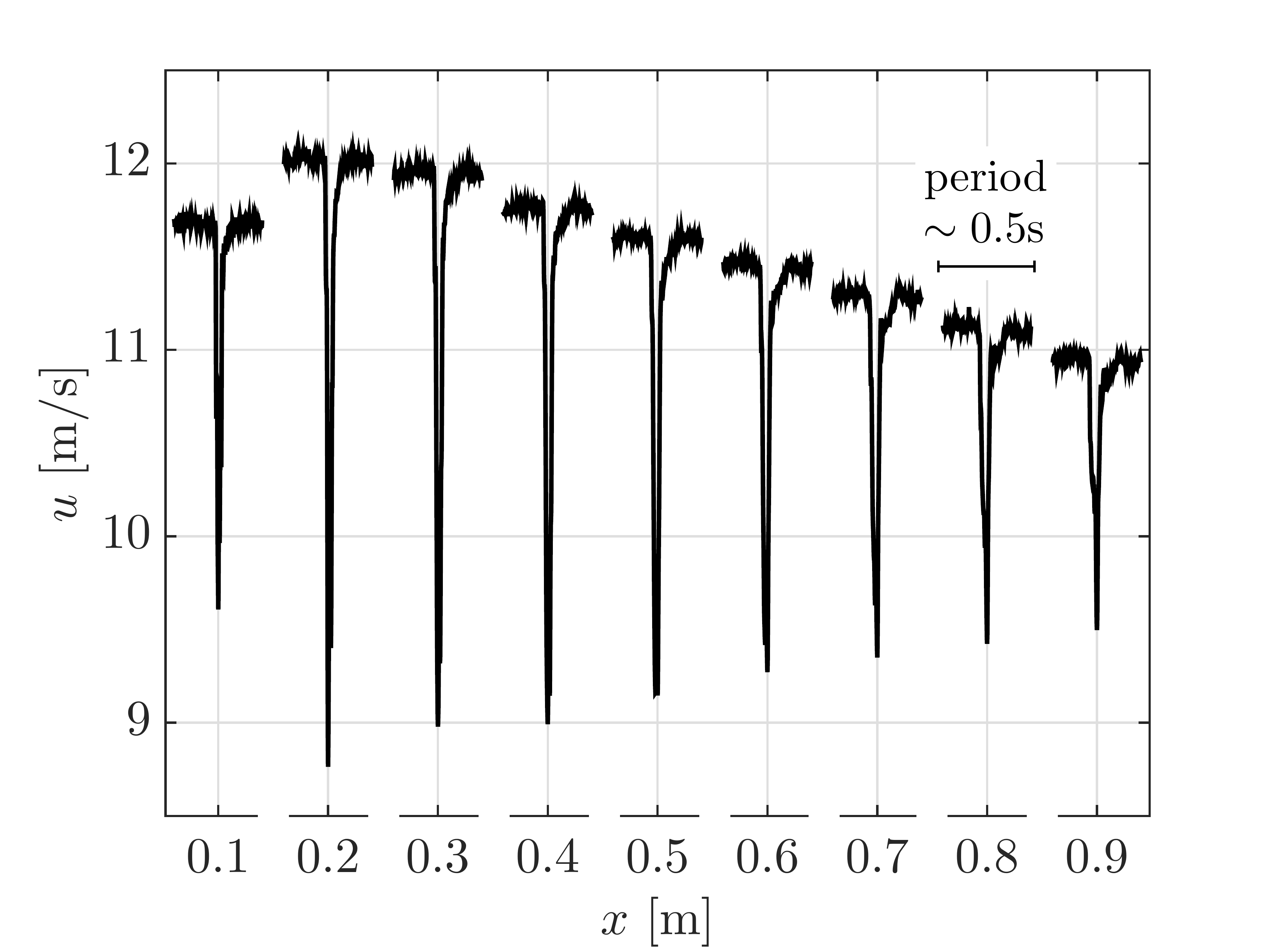}
\vspace{0.5cm}
 \caption{Velocity modulations at different downstream positions }
\label{fig:modulations_downstream}
\end{figure}

\begin{figure}
\centering  
\includegraphics[width=0.6\textwidth]{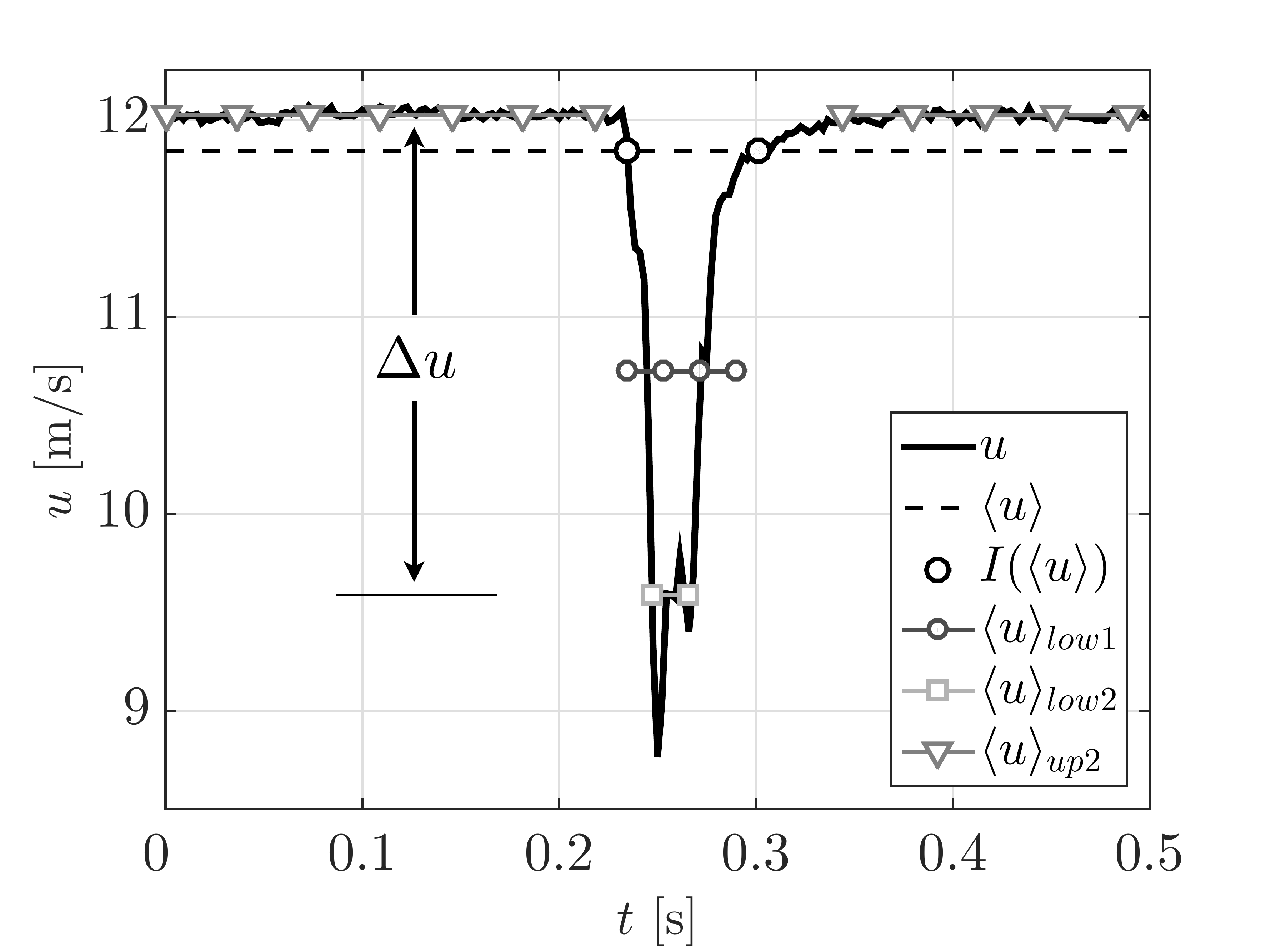}
\vspace{1cm}
\caption{Single exemplary velocity modulation over time and illustration of procedure to estimate  $\Delta u$}
\label{fig:one_modulation}
\end{figure}

\begin{figure}
\centering
a)\includegraphics[width=0.45\textwidth]{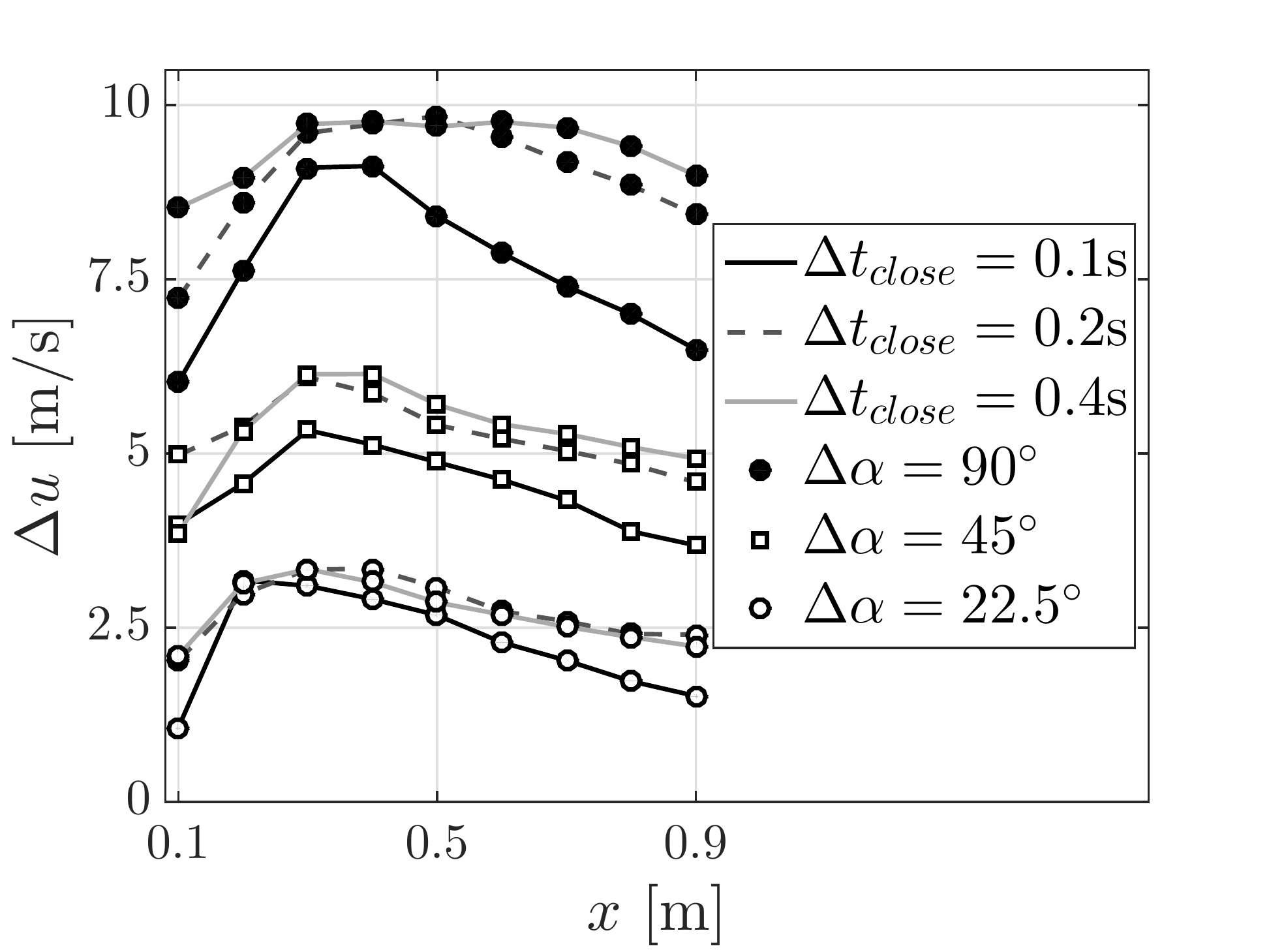}
b)\includegraphics[width=0.45\textwidth]{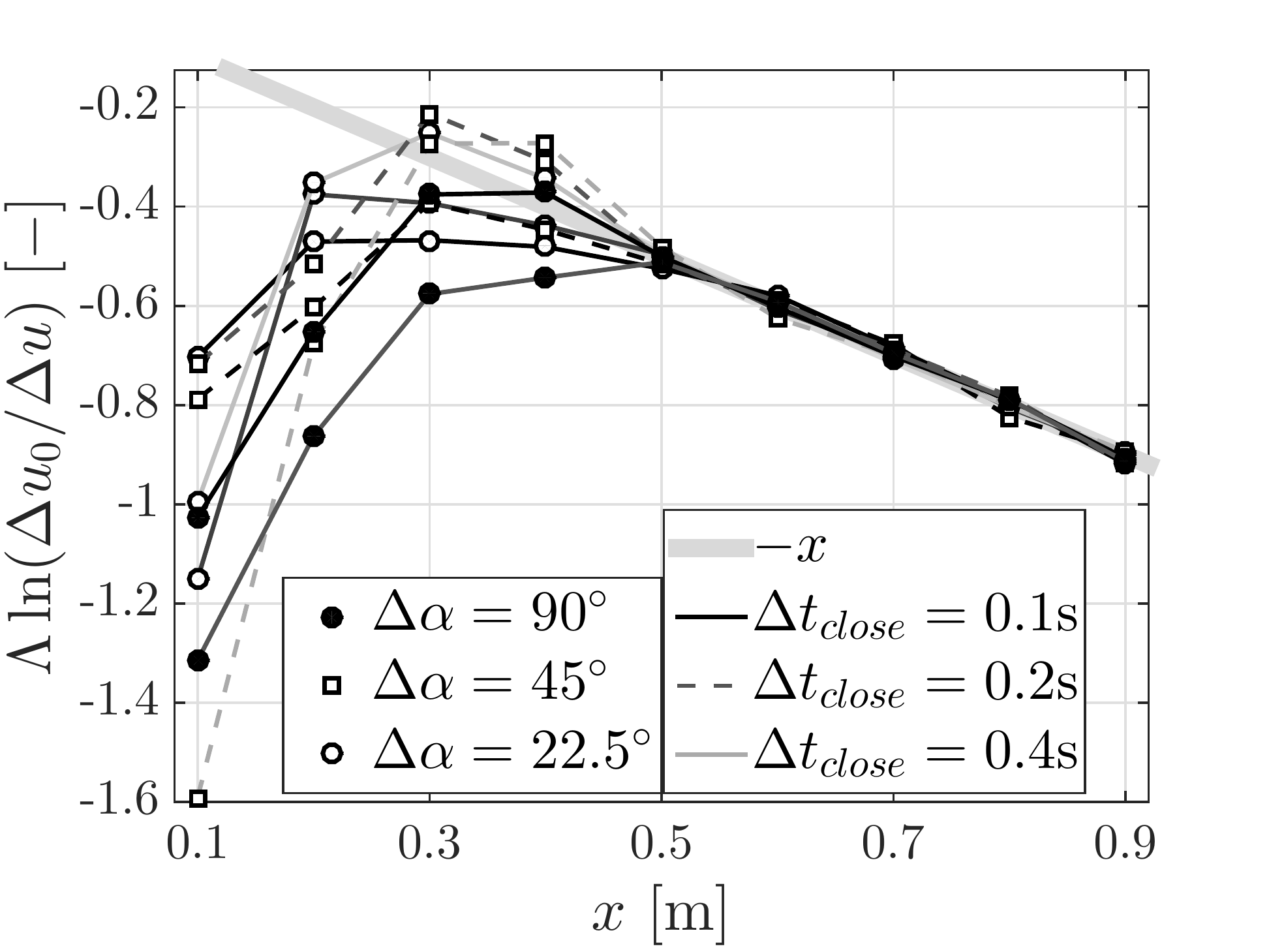}
\vspace{1cm}
 \caption{Comparison of various flow modulations generated under the conditions  summarized in table \ref{table:exp1}.
Figure a) shows the modulation depth $\Delta u$ over downstream position $x$ and b) shows the result of exponential approach, see eq. (\ref{eq:app_decay})}
\label{fig:modulations}
\end{figure}

\begin{figure}
  \centering 
a)\includegraphics[width=0.45\textwidth]{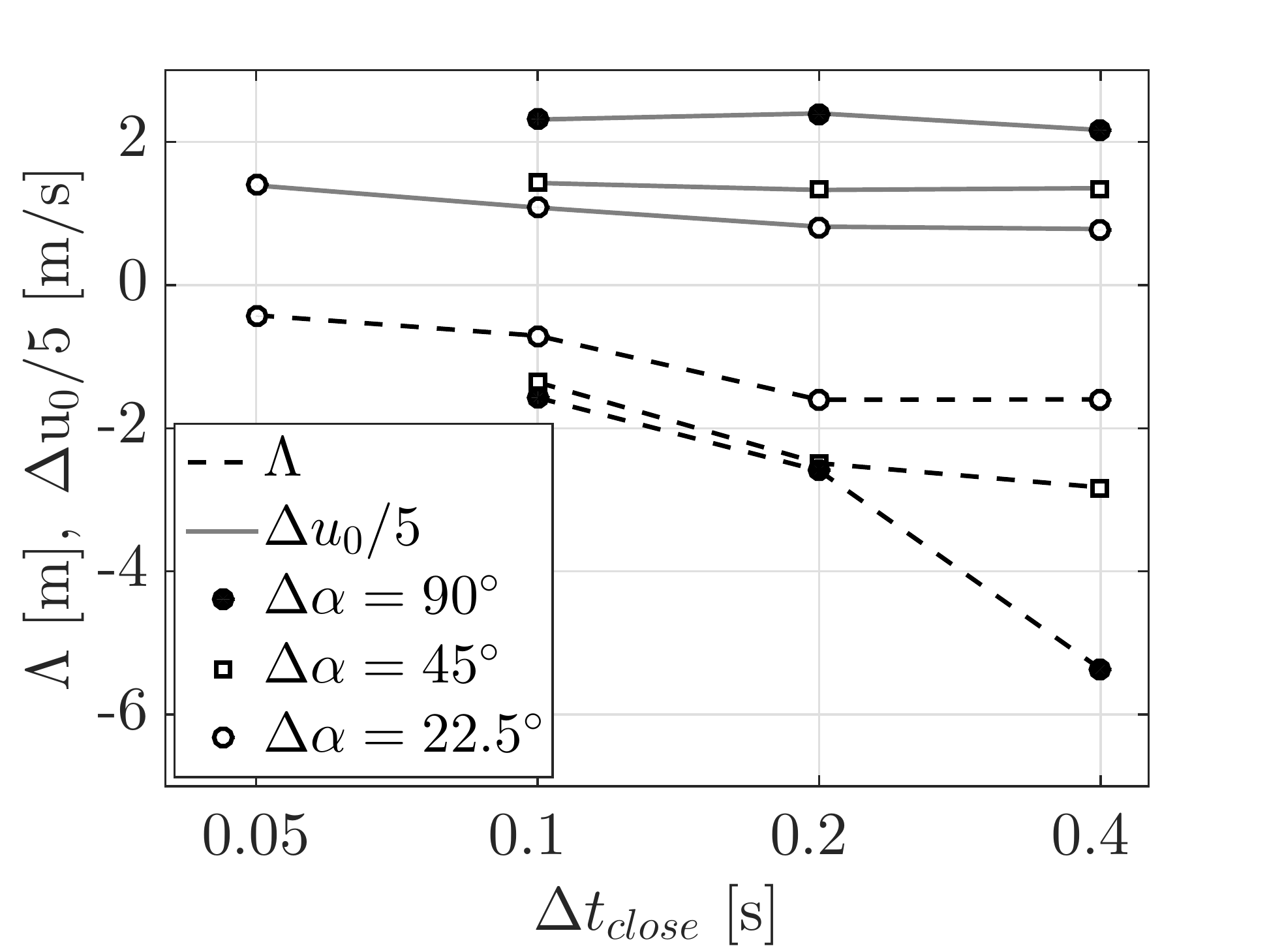}
b)\includegraphics[width=0.45\textwidth]{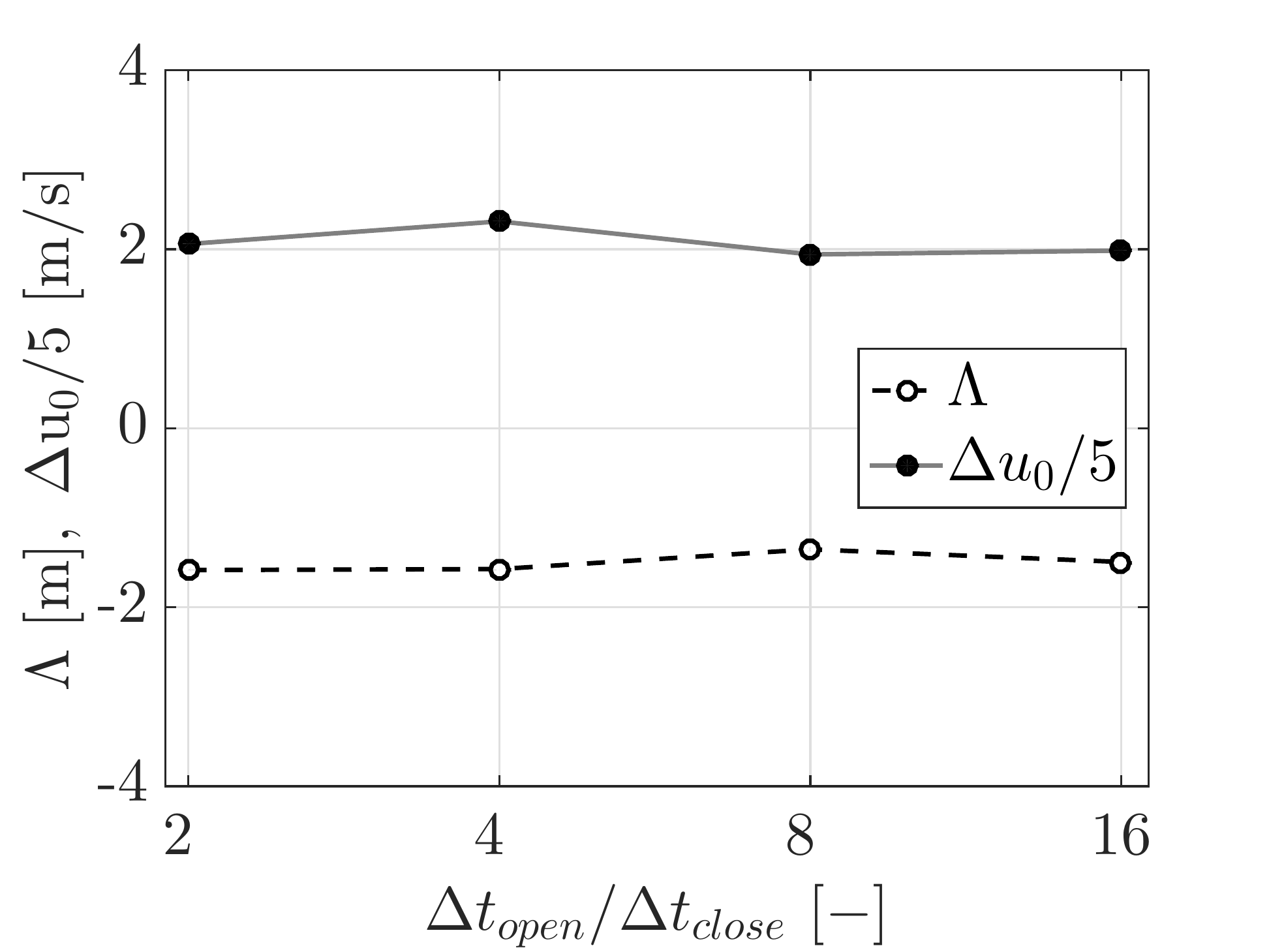}
\vspace{0.5cm}
\caption{Decay of various flow modulations presented in terms of $\Lambda$ and $\Delta u_0$, see eq. (\ref{eq:app_decay}). Results related to the decay of modulations are shown in figure a), experiment (i). Figure b) shows results to the interaction of flow modulations, experiment (ii)}
\label{fig:A_B_parameter}
\end{figure}

To understand dynamics of flow modulations on their way downstream, two experiments are performed. 
A first experiment (i) is dedicated to the decay of flow modulations and a second experiment  (ii) to interactions of  modulations.
These two investigations are based on simple and well-defined motions of the flaps (according to the constant blockage configuration), which generate specific flow modulations. 
Figure \ref{fig:Pulse_AoA} exemplarily presents  such well-defined motion of the inner flaps.
The motion is periodically and consists mainly of two flap positions, an open and a closed position, $\alpha_{open}$ and $\alpha_{close}$, with corresponding time intervals $\Delta t_{open}$ and $\Delta t_{close}$. 
It results in flow modulations, where the velocity is high for $\Delta t_{open}$ and reduced for $\Delta t_{close}$.
An intensity measure of a flow modulation can be defined by $\Delta \alpha=\alpha_{close} - \alpha_{open}$.
The purpose of these explicit flow modulations is that they are simple to characterise, since only $\alpha_{open}$, $\alpha_{close}$, $\Delta t_{open}$ and $\Delta t_{close}$ determine the modulation.
In case $\alpha_{open}=0^\circ$ and $\Delta t_{close} \ll \Delta t_{open}$, we expected that $\Delta \alpha$ and $\Delta t_{close}$ characterise the flow modulations. This simplified case is used to investigate decay features of modulations, experiment (i).
A variation of \mbox{experiment (i)} is the case where the ration $\frac{\Delta t_{open}}{\Delta t_{close}}$ is varied to investigate interactions of modulations, which is investigated in \mbox{experiment (ii)}.
\\
To give an impression of the flow modulations shown in figure \ref{fig:Pulse_AoA}, figure \ref{fig:modulations_downstream} shows resulting velocity modulations at different downstream positions $x$. 
To highlight the velocity modulations, the measured velocity signals are smoothed with a moving average and an averaging over many periods at a fixed downstream positions. 
This smoothing procedure corresponds to an averaging over $\approx 180$ samples or a time interval of $\approx$~0.009~s.
\\
Considering one modulation, a high speed plateau, which corresponds to $\Delta t_{open}$ and a velocity drop, which corresponds to $\Delta t_{close}$ can be recognized.
It can be seen that the velocity modulations evolve downstream.
This evolution can be characterised by the velocity difference $\Delta u$ between the plateau and bottom of the modulation as function of the downstream position. 
Below, the issue of decay and interaction is studied with $\Delta u(x)$.
\\
Next, we explain how $\Delta u$ gets estimated. 
Therefore, a two step iterative procedure gets introduced, which is  reliable in estimating $\Delta u$ for various kinds of modulations.
The procedure is illustrated in figure \ref{fig:one_modulation}, where a 
 zoom of single velocity modulation ($\sim$one period) is shown. 
This velocity sequence consists basically of three parts, an unmodulated (plateau), a modulated ($\sim$bottom) and again an unmodulated part (plateau). These parts get 
detected by the procedure.
Therefore the following steps are done: 
Firstly, the mean velocity $\langle u \rangle$ is determined. 
This mean has two intersections $I(\langle u \rangle)$ with the modulated flow $u$. 
Secondly, between these intersections the mean of the velocity signal is determined, which leads to $\langle u \rangle_{low1}$.  $\langle u \rangle_{low1}$ has two new lower intersections with the velocity signal. 
Thirdly, the mean between these new intersections gives the bottom velocity $\langle u \rangle_{low2}$. 
Fourthly, the corresponding approach (means \textit{outside} the intersections) is done to determine first $\langle u \rangle_{up1}$ and then $\langle u \rangle_{up2}$ as shown in fig. \ref{fig:Pulse_AoA}.
Finally, the difference between $\langle u \rangle_{up2}$ and $\langle u \rangle_{low2}$ is defined as velocity drop $\Delta u$. 
\\
In  experiment (i) various flow modulations are generated by systematically changing $\Delta \alpha$ and $\Delta t_{close}$. Resulting flows are measured in different wake positions $x$. Details are summarized in table \ref{table:exp1}. 
Figure \ref{fig:modulations} a) shows how
$\Delta u$ develops over downstream position.
All modulation depths evolve downstream, first the depth increases and then it decreases. 
Note that a similar evolution of turbulence intensity and intermittency is found and discussed in \cite{Reinke2013}.
The influence of $\Delta \alpha$ appears in the magnitude of the drop, where a high $\Delta \alpha$ leads to a high $\Delta u$. 
\begin{table} 
\centering
 \begin{tabular}{  c  |ccc} 
 $\Delta \alpha \ [^\circ]$ (with $ \alpha_{open} = 0^\circ $)    	& $\Delta t_{close} \ [\rm{s}]$ (with $\frac{\Delta t_{open}}{\Delta t_{close}}$=4)   \\
  \hline
      22.5, 45, 90                                  				&     0.1, 0.2, 0.4                              \\
 open circle, square, solid circle         				&    bright grey,  dashed, black         \\                                 
\end{tabular}
\vspace{0.5cm}
\caption{Summary of investigated parameters in experiment (i) and their corresponding line coding in figure~\ref{fig:modulations}}
\label{table:exp1}
\end{table} 
\\
To quantify the decrease of $\Delta u$ in more detail, an exponential fitting,
\begin{eqnarray}
\Delta u (x) &\mathrel{\widehat{=}}& \Delta u_0 \ e^{{x}/{\Lambda}} 
\label{eq:app_decay}\\
\Leftrightarrow \hspace{1cm} -x &=& \Lambda \ln \left( \frac{\Delta u_0}{\Delta u(x)} \right). 
\label{eq:app_decay2}
\end{eqnarray}
is applied on every $\Delta u$ evolution
in a range of \mbox{0.5~m$\le x \le$0.9~m}. In this range all $\Delta u$ decrease, 
(besides the most stable $\Delta u$ evolution (top curve in figure \ref{fig:modulations} a)).
$\Lambda$ and $\Delta u_0$ are fit parameters. $-\Lambda$ can be related to a relaxation length. Note that $\Delta u_0$ is not the depth of modulation at $x$=0, but can be interpreted as \textit{virtual origin}, as commonly done in investigations regarding the decay of grid generated turbulence, e.g. 
\cite{Mohamed1990,Skrbek2000}.
To check the validity of the exponential approach, eq. (\ref{eq:app_decay}) is transposed to eq. (\ref{eq:app_decay2}) and in
figure \ref{fig:modulations} b) the result of the transformation is shown, which nicely confirms the proposed exponential decay.
Thus, $\Lambda$ as well as $\Delta u_0$ are proper parameters to discuss the decay characteristics of the modulations.
\\
It is worth noting
that exponential and power law features are quite common for turbulent flows.
There is an ongoing discussion for grid generated turbulence whether or not its turbulence decay obeys an exponential or a power law, cf. \cite{Mydlarski1996,George2009,Sinhuber2015}. 
An exponential decay of flow modulations is predicted theoretically by \cite{Kraichnan1964} and is experimental verified by \cite{Camussi1997}, 
under the conditions of homogenous isotropic turbulence.
However, this idealised case is not expected for the considered flows of the experiments presented here, cf. \cite{Makita1991,Poorte2002,Kang2003,Reinke2013}.
A third example of exponential features are exponentially distributed lifetimes of flow modulations, which are known from the start of turbulence, cf. \cite{Hof2006,Barkley2011}. 
\\
Figure \ref{fig:A_B_parameter} a) shows the decay of modulations  in terms of fit parameters $\Lambda$ and $\Delta u_0$, as function of the modulation length $\Delta t_{close}$ and modulation intensity $\Delta \alpha$. 
The parameter $\Lambda$ describes how fast the decay takes place. Thus, a large negative $\Lambda$ corresponds to a slow decay and a longer relaxation length. 
Since $\Lambda$ deviates more from zero the more $\Delta t_{close}$ increases, 
figure \ref{fig:A_B_parameter} a) shows that longer modulations are more stable downstream. 
Furthermore, a tendency is shown that stronger modulations in terms of $\Delta \alpha$ are more stable downstream. 
\\
Figure \ref{fig:A_B_parameter} a) shows also $\Delta u_0$, which gets bigger the more $\Delta \alpha$ increases and hence  confirms the observation in figure \ref{fig:modulations} a).
The three $\Delta u_0$ curves are basically parallel to each other, therefore a rather weak dependency of $\Delta u_0$ on modulation duration $\Delta t_{close}$ can be assumed. 
The slight increase of $\Delta u_0$ to shorter times ($\Delta t_{close}<0.2$~s and $\Delta \alpha=22.5^\circ$) might be related to dynamic stall effects on the grid flaps.
\\
In experiment (ii) various periodic flow modulations are generated with varying ratios of $\frac{\Delta t_{open}}{\Delta t_{close}}$.
Table \ref{table:exp2} summarizes the investigated parameters.
The investigation is based on the idea that in the case of interactions of modulations
decay features 
change and hence the decay depends on the ratio $\frac{\Delta t_{open}}{\Delta t_{close}}$.
Figure \ref{fig:A_B_parameter} b) shows that $\Lambda$ and $\Delta u_0$ show no significant dependence in the investigate range of $\frac{\Delta t_{open}}{\Delta t_{close}}$.
This result might be due to the fact that
the modulation lengths are quite long compared to the measurement position, i.e. \mbox{$\Delta t_{close}=0.1$~s} and $\langle u \rangle=10~\frac{\rm{m}}{\rm{s}}$ corresponds to 1~m modulation length.
\begin{table}
\centering
 \begin{tabular}{c|c|ccccc} 
$\Delta \alpha \ [^\circ]$&   $\Delta t_{close} \ [\rm{s}]$  &$\frac{\Delta t_{open}}{\Delta t_{close}}$        $[-]$  \\
\hline
    90 &                    0.1 &   2, 4, 8,  16                  \\
\end{tabular}
\vspace{0.5cm}
\caption{Summary of investigated parameters in experiment (ii)}
\label{table:exp2}
\end{table}
\\
We would like to close this section by two comments.
First, in case of other configurations of axes (e.g. no constant blockage configuration) the resulting flow features of modulations might vary.
Second, flow modulations are investigated with common active grid motion parameters. 
For shorter modulations, we expect interaction of modulations,
e.g. likewise to a  K\'arm\'an street cf. \cite{Mandal2011}.

\section{Examples of active grid wake estimation}
\label{sec:Examples}
This section deals with three examples which explain the generation of synthetic velocity time series $u_{syn}$ and the estimation of active grid wake features.
Note that synthetic velocity time series are composed on a computer and these $u_{syn}$ are based on reference data, for instance reference velocity time series $u_{ref}$. 
Synthetic velocity time series are validated with time series $u_{wt}$, which are measured in the wind tunnel and with statistical features of $u_{wt}$. 
\\
The estimating approach is based on the results obtained in section \ref{sec:Method}.
Thereby, the dynamic features of interactions are rated as weak and also the dynamics of decay are rated as weak for \textit{long} modulations. 
Therefore, the approach is based on the static wake calibration (sec. \ref{sec:Static}) and neglects dynamics in a first approximation.
\\
The first example  (sec. \ref{sec:Examples_1})  takes up the commonly used random motion protocols, e.g. \cite{Makita1991,Mydlarski1996,Poorte2002,Cekli2010,Knebel2011,Thormann2014,Hearst2015} and shows to a given 
random motion protocol
a corresponding wake estimation.
\\
The second example  (sec. \ref{sec:Examples_2})  is dedicated to the inverse problem of the first example.
Here, a flow feature, namely a probability density function of velocity, is given as a reference and a motion protocol which generates this feature is sought.  
\\
The third example  (sec. \ref{sec:Examples_3})  is a paradigm for a reproduction of a reference time series.
Thus, it shows how the wind tunnel flow can be modulated so that the modulated flow matches a given reference velocity time series. 
Thereby, measured wind conditions of the atmospheric boundary get reproduced in the wind tunnel.


\subsection{Example: random motion}
\label{sec:Examples_1}
\begin{figure}
\centering
\includegraphics[width=0.6\textwidth]{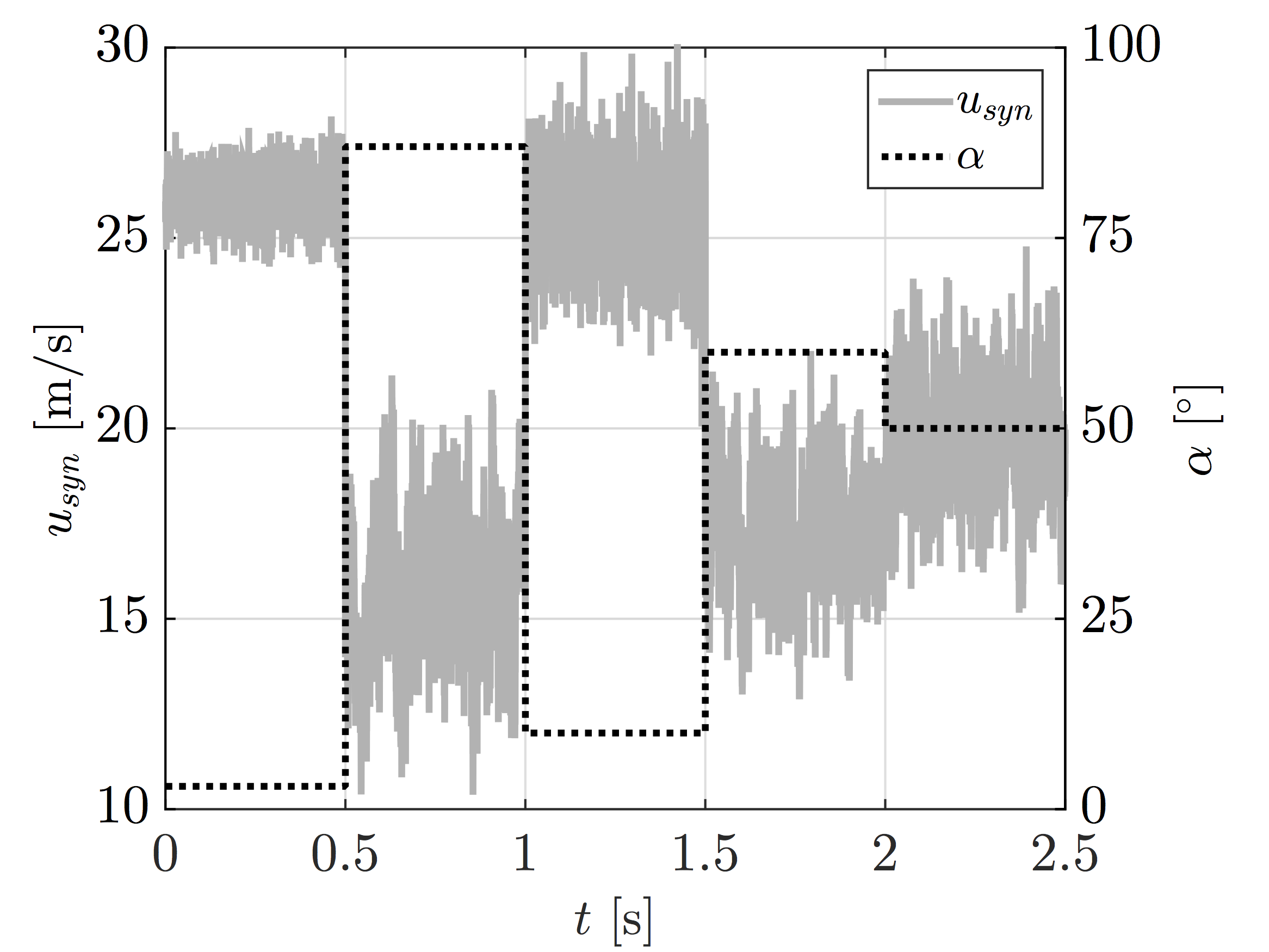}
\vspace{0.5cm} \caption{Synthetic velocity time series $u_{syn}$ according to a given reference $\alpha$ time series}
\label{fig:AG_Chara_von_AoA_zu_u}
\end{figure}

\begin{figure}
\centering
a)\includegraphics[width=0.6\textwidth]{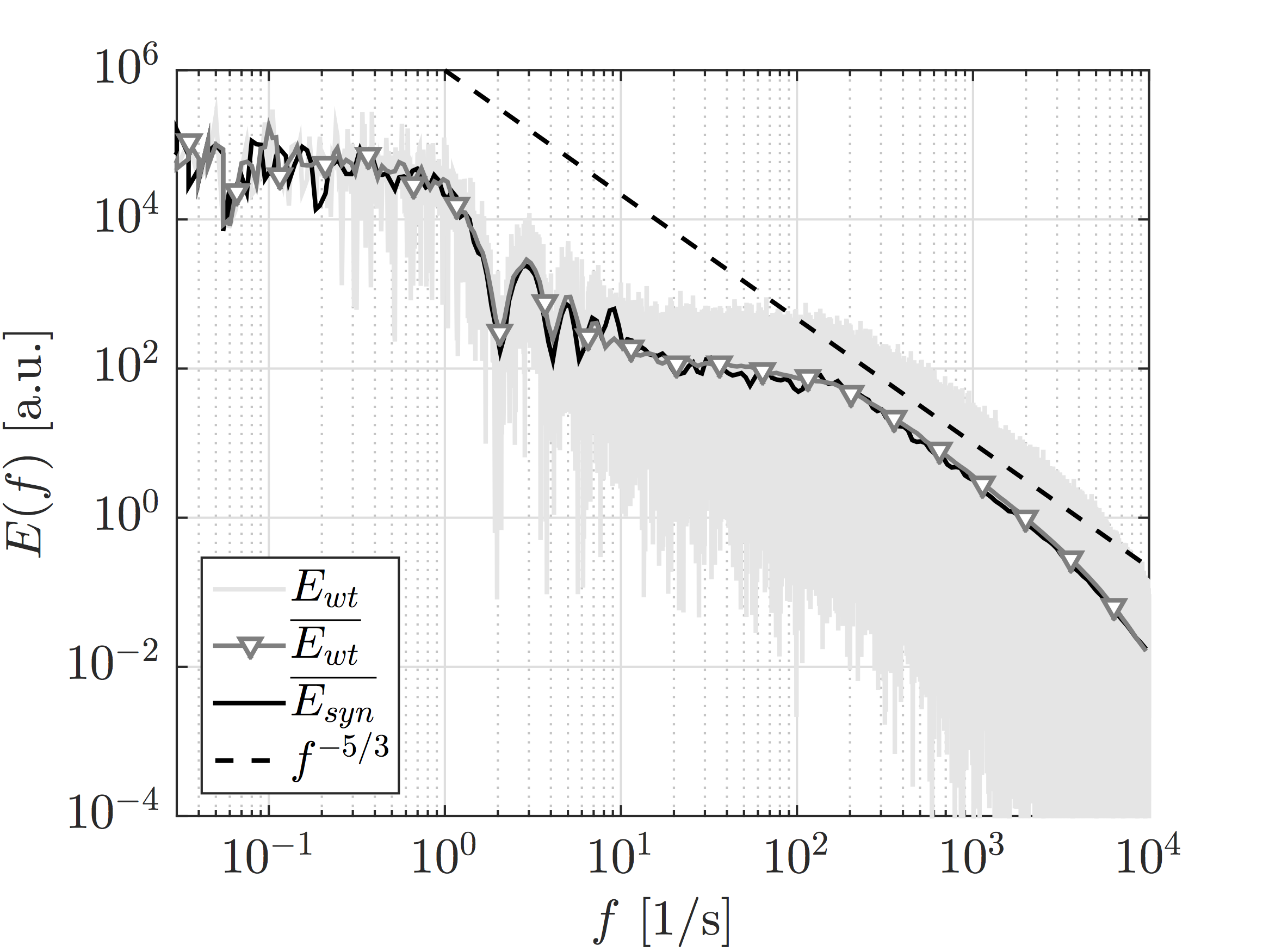}
b)\includegraphics[width=0.6\textwidth]{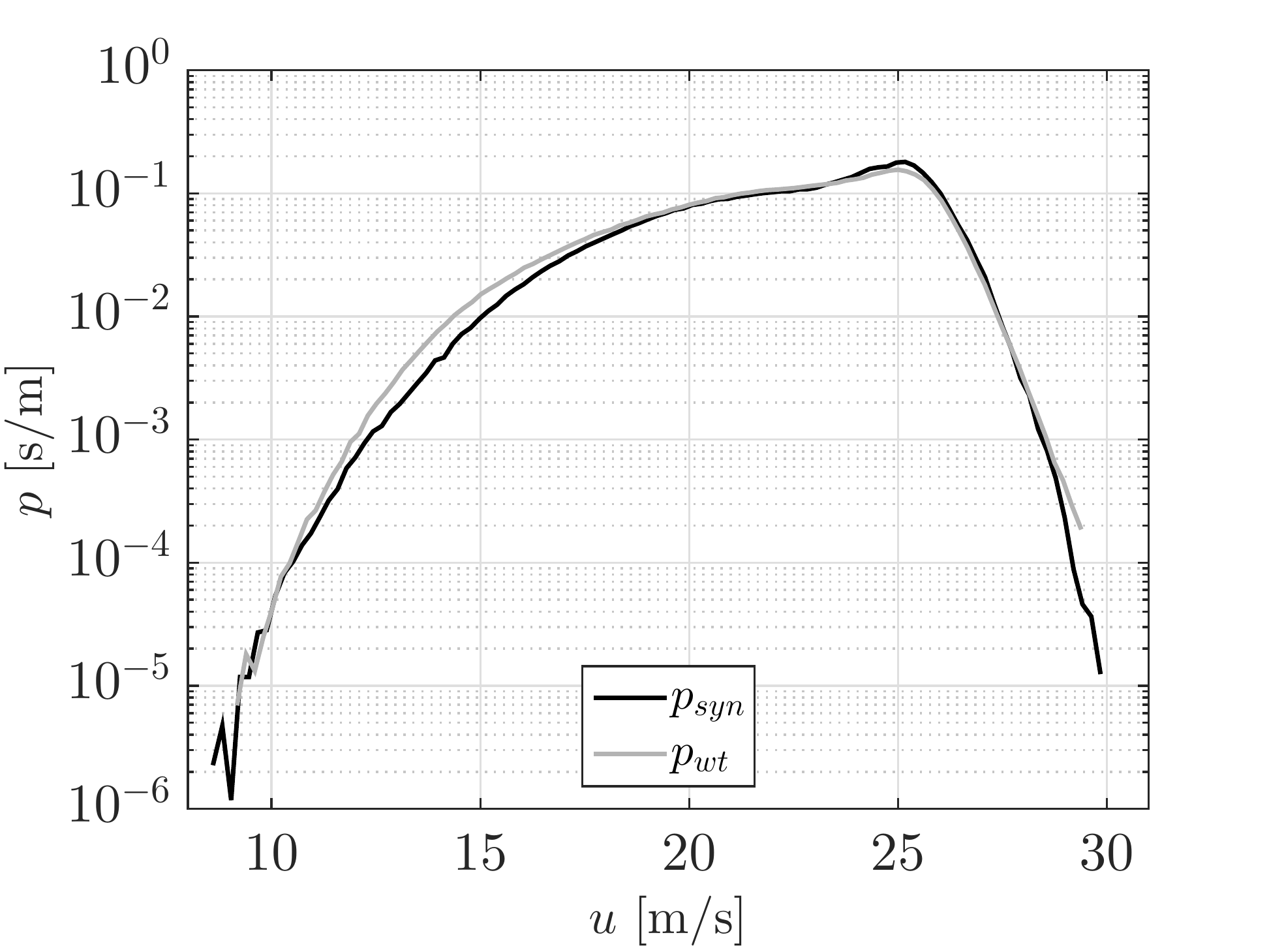}
c)\includegraphics[width=0.6\textwidth]{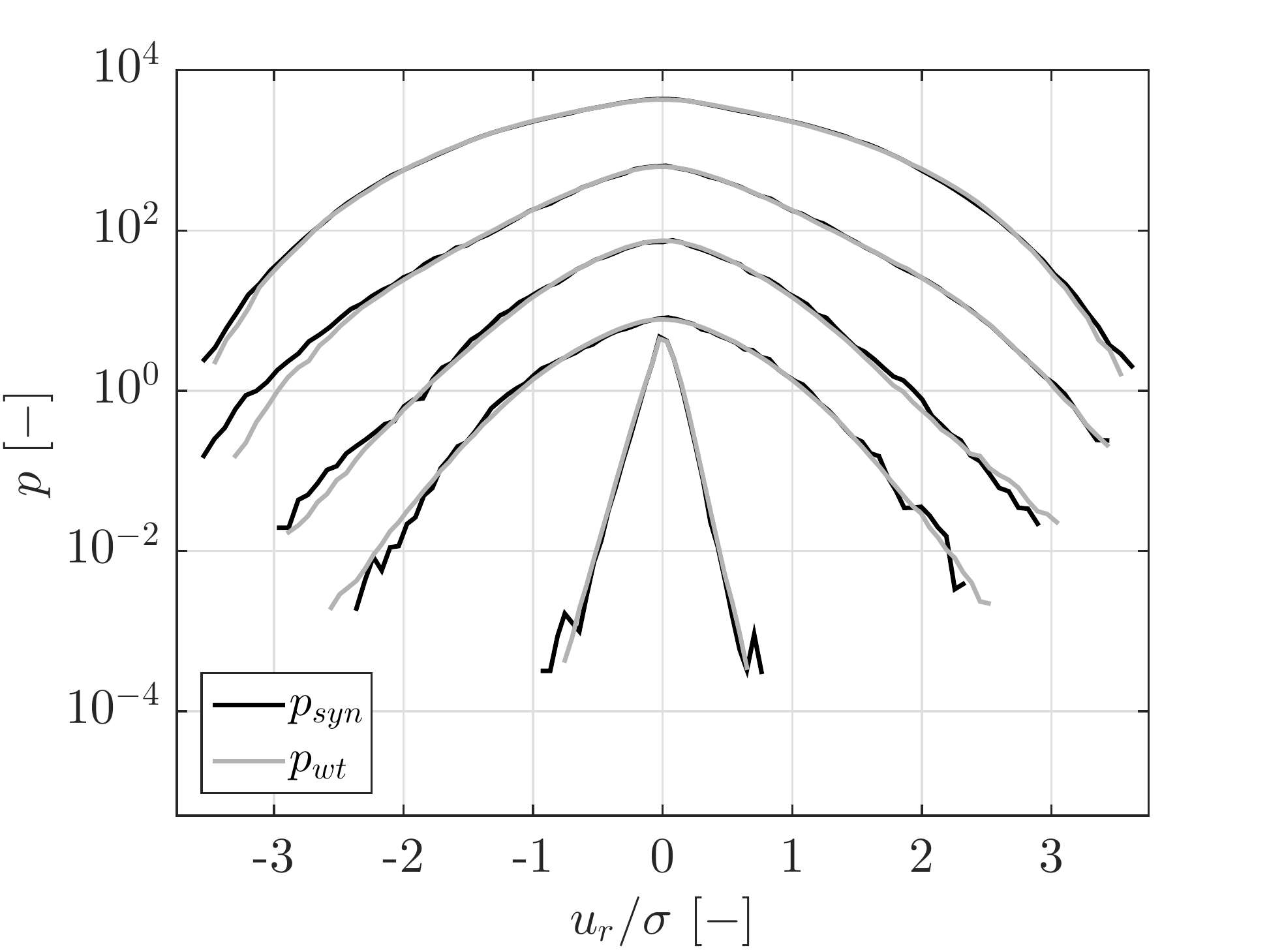}
\vspace{0.5cm} \caption{Comparison of a synthetic time series with data from a wind tunnel experiment, both generated under the same flow conditions. These are compared in terms of (a) power spectral density $E(f)$, (b) probability density function of velocity and (c) probability density functions of velocity increments}
\label{fig:example_1}
\end{figure}

This example consists of two parts.
First, an simplified example illustrates a procedure of building up a synthetic velocity time series.
Second, such a synthetic time series is compared with measured data.
\\
The simplified example is illustrated in figure \ref{fig:AG_Chara_von_AoA_zu_u}. It presents an given artificial reference $\alpha_{ref}$ time series, which represents the motion of active grid flaps. The various angular positions $\alpha$ stay constant for \mbox{$\Delta t = 0.5$~s.} 
In order to generate a synthetic velocity time series to this $\alpha$ time series the following procedure has to be made with the wake calibration, fig. \ref{fig:AG_Chara_wake} and the $\alpha$ time series, fig. \ref{fig:AG_Chara_von_AoA_zu_u}:
One searches to the first flaps position with $\alpha\approx3^\circ$ in fig. \ref{fig:AG_Chara_von_AoA_zu_u}, the corresponding velocity fluctuation in fig.~\ref{fig:AG_Chara_wake}, cut out around this fluctuation a piece of fluctuations with the required time length of $\Delta t$. 
Add to this piece of fluctuations the corresponding trend velocity, fig. \ref{fig:AG_Chara_wake} and the first 0.5~s of the so-called synthetic velocity time series $u_{syn}$ is generated in figure \ref{fig:AG_Chara_von_AoA_zu_u}.
This procedure gets continued for all reference $\alpha$-values and the complete synthetic velocity time series is generated by sticking all pieces in a series, as shown in figure \ref{fig:AG_Chara_von_AoA_zu_u}. 
\\
In this approach the transition motion to the next $\alpha$ is neglected, since transition time is very short. 
This neglect is also reasonable from an energetic perspective. The maximal rotation velocity at the flaps tips is $u_{rot}\approx0.83~\frac{\rm{m}}{\rm{s}}$, which is quite slow compared to the inflow velocity $u_{\infty}\sim10~\frac{\rm{m}}{\rm{s}}$, or in terms of kinematic energy $(u_{\infty}/u_{rot})^2\approx 144$. 
\\
Finally, the whole synthetic velocity time series is low pass filtered with the half of the sampling frequency of the wake calibration measurement, here $f_l=10$~kHz. 
This filtering reduces artificial frequencies, which get in $u_{syn}$ due to the process of sticking together different fluctuation pieces.
\\
For testing the estimating approach  $u_{syn}$ is compared to a measured velocity times series $u_{wt}$. Both time series belong to a random motion pattern, which is defined as
$\alpha_{ref}(t) = \langle \alpha \rangle + \omega(t)$ with $\langle \alpha \rangle =0^\circ$ and $\omega(t)$  
is a random number. 
The random numbers are Gaussian distributed with standard deviation of $\sigma(\alpha)=\sqrt{\langle \omega(t)^2 \rangle}=11^\circ$. 
Every 0.5~s the flaps move to a new $\alpha$, likewise to figure \ref{fig:AG_Chara_von_AoA_zu_u}. 
Thus, $\langle \alpha \rangle$ and $\sigma(\alpha)$ specify the flow modulation.
\\
Figure \ref{fig:example_1} shows a stochastic comparison of $u_{syn}$ and $u_{wt}$. 
Also the following quantities are subscripted ${syn}$ and ${wt}$ and belong likewise to the synthetic time series and the wind tunnel measurement. 
Figure \ref{fig:example_1} a) presents the power spectral density $E(f)$ over frequency $f$, as well as its window averaged spectra $\overline{E}$. 
To facilitate the comparison, an averaging is done in frequency space
with an exponential increase of window size.
Thus high frequencies get stronger averaged than lower frequencies.
Figure \ref{fig:example_1} b) presents probability density function (PDF) of velocity $u$. 
Figure \ref{fig:example_1} c) presents PDFs of velocity increments, which is a comparison in scale. 
Velocity increments are defined as  \mbox{$u_r=u(x+r)-u(x)$} (normalized by $\sigma=\sqrt{\langle u^2 \rangle}$), presented scales are 
$r = 0.5$~mm,  3.2~cm,  37~cm, 2~m and 7.8~m (from bottom to top). Spatial differences are estimated by applying Taylor's hypotheses.  
Note that PDF bins with less than 50 events are not considered in figure \ref{fig:example_1} b) and c) and for reason of clarity the PDFs in figure \ref{fig:example_1} c) are shifted vertically.
\\
All in all, the stochastic features of the synthetic time series $u_{syn}$ are in very good agreement with the experimental time series $u_{wt}$
Thus, we conclude that the approach is well-suited for such a wake estimation.  
\\
We would like to give a comment on the shape of $E(f)$, figure \ref{fig:example_1} a), which is different to the common scaling, $\sim f ^{-5/3}$,  
cf. \cite{kolmogorov1941b,Obukhov1941a,Obukhov1941b} 
mentioned in \cite[p. 98]{Frisch2001}. The reason for this particular shape is the specific active grid motion protocol $\alpha(t)$, which is not adapted for the generation of homogeneous isotropic turbulence.
\\
Note that this procedure of generating synthetic velocity time series  
from a given $\alpha$ time series is mainly helpful for two aspects. 
First, it shows what kind of velocity signal can be expected, e.g. in terms of its statistic features. 
Second, it is possible to study the impact of different control strategies and motion patterns. 
Therefore, extensive parameter studies can be done on a computer, instead of wind tunnel experiments. This procedure is further explained in the next example (sec. \ref{sec:Examples_2}).

\subsection{Example: motion according to a stochastic feature}
\label{sec:Optimization}
 \label{sec:Examples_2}
\begin{figure}
\centering
a)\includegraphics[width=0.45\textwidth]{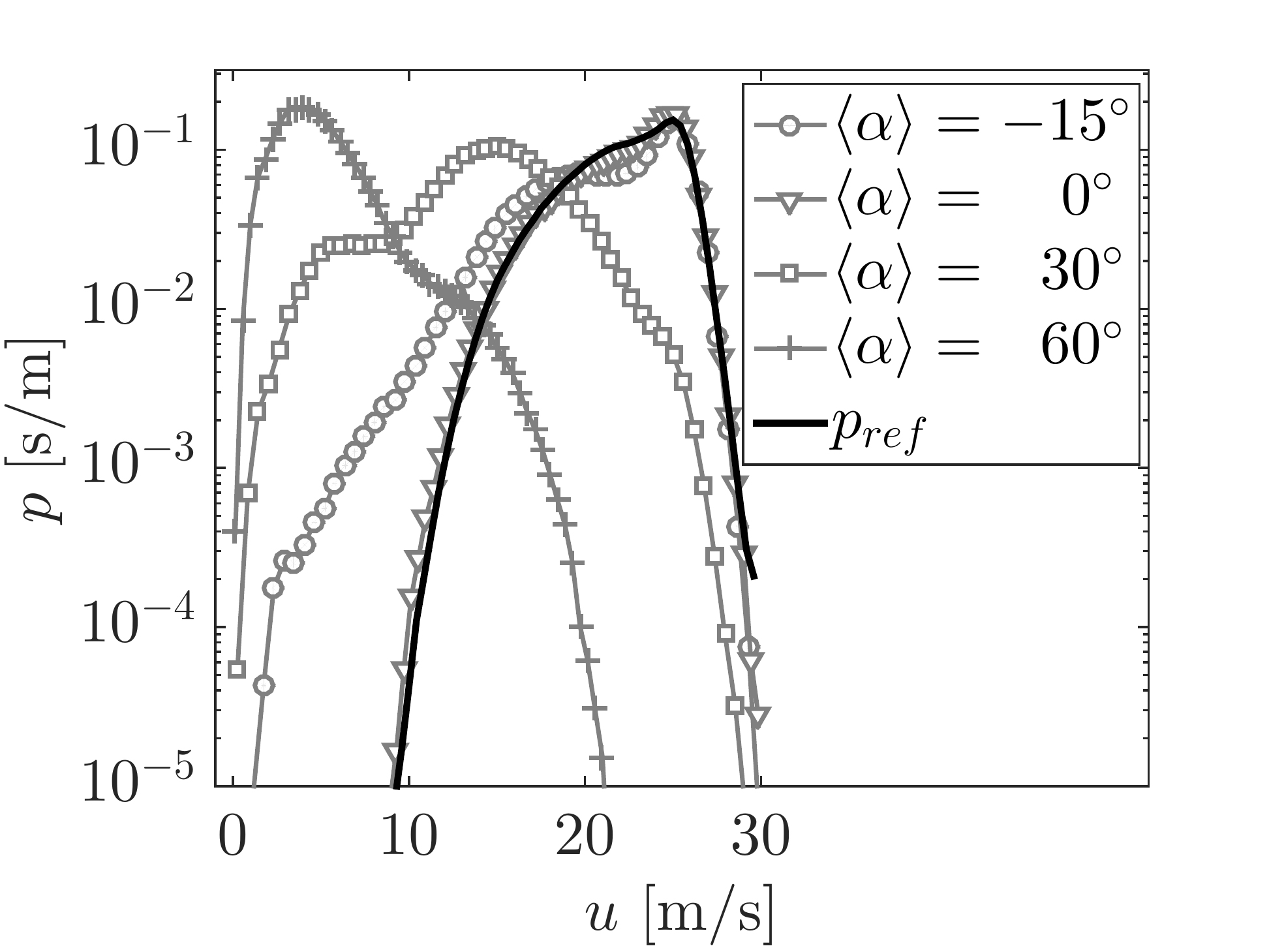}
b)\includegraphics[width=0.45\textwidth]{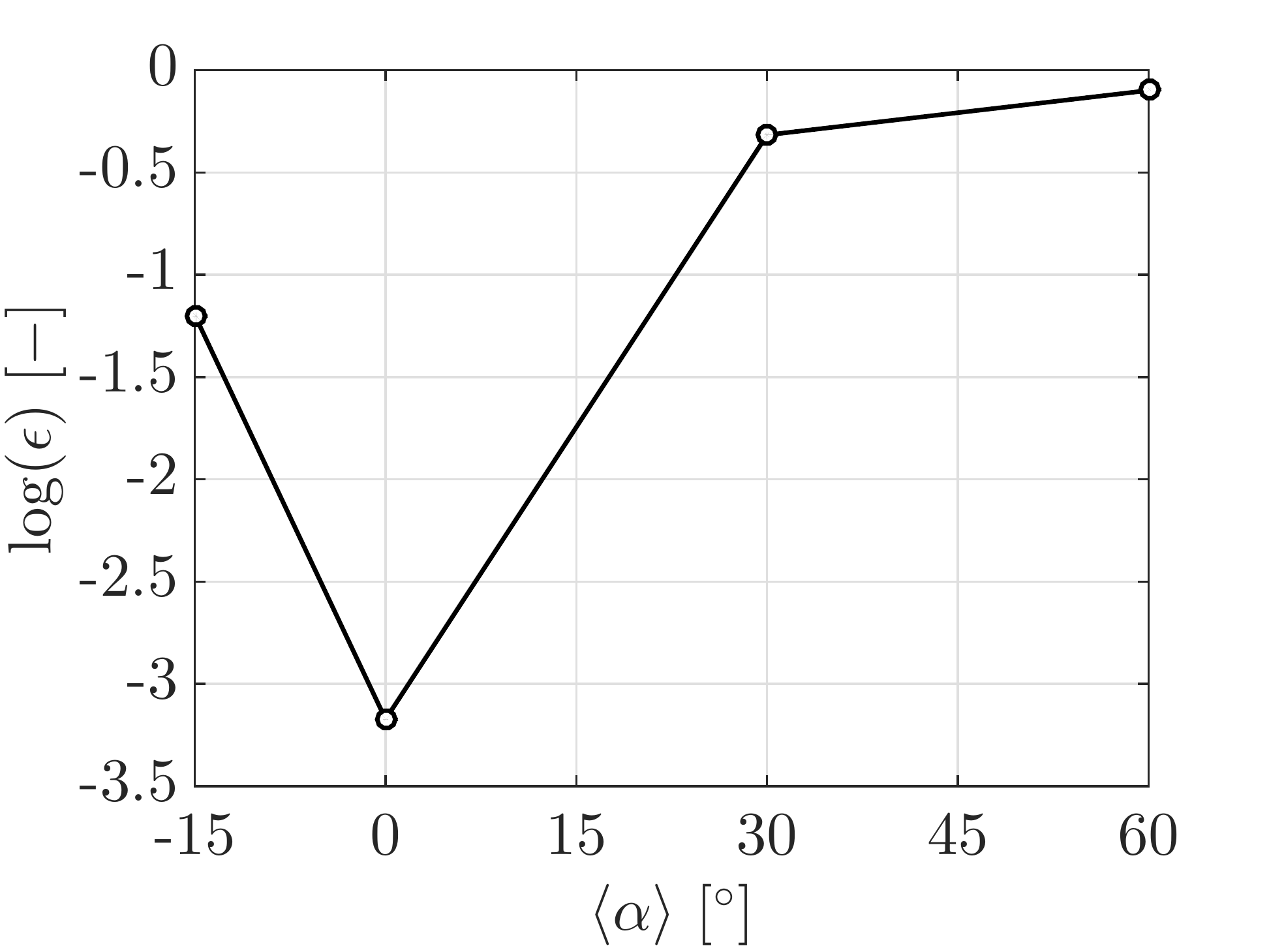}
 \vspace{0.5cm} \caption{a) Comparison of reference PDF with various synthetic PDFs. b) Conformity of synthetic PDFs with reference PDF in terms of $\log(\epsilon)$}
\label{fig:example2PDF}
\end{figure}

\begin{figure}
\centering
\includegraphics[width=0.6\textwidth]{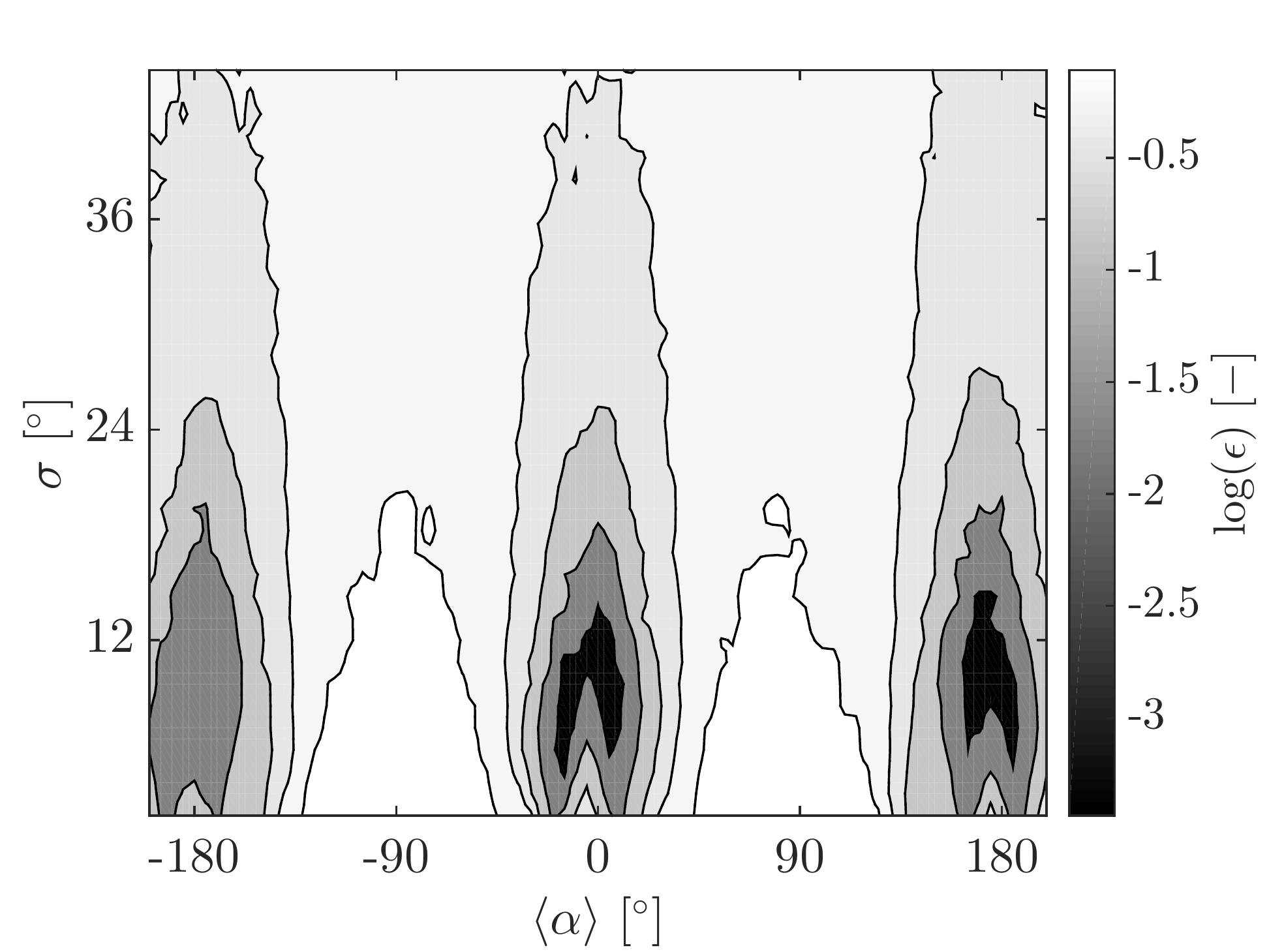}
 \vspace{0.5cm} \caption{ Illustration of the parameter study, 
 log($\epsilon$) as function of motion parameters $\langle \alpha \rangle$ and $\sigma$, which shows how well $p_{ref}$ and $p_{syn}$ are in agreement and the corresponding motion parameters 
}
\label{fig:example2epsilon}
\end{figure}

In this example  
a wake feature is given and a corresponding motion pattern is worked out.
Therefore, a parameter study with synthetic time series is done.
The resulting parameter set corresponds to a proper flow modulation.
Without deeper meaning, we selected a  probability density function of velocity as wake feature, which we want to reproduce and it is defined as reference $p_{ref}$.
In the interest of simplification, the velocity PDF $p_{wt}$ is used as reference, shown in figure \ref{fig:example_1} b), thus $p_{ref}=p_{wt}$.
Hence, the aim of this example is to find a flaps motion pattern, which generates a flow with $p_{ref}$.
\\
First, one has to select a certain motion pattern, which obeys certain motion parameters. 
In this example we know already the motion pattern from section \ref{sec:Examples_1}. Thus, the parameter study results in the proper values of $\langle \alpha \rangle$ and $\sigma(\alpha)$.
The issue of guessing a proper motion pattern and parameters is commented at the end of this section.
\\
Next, various synthetic velocity time series $u_{syn}$ get generated according to a systematic change of the motion parameters. 
 Afterwards,  PDFs of velocity are determined from these $u_{syn}$.
Figure \ref{fig:example2PDF} a) shows a few exemplary PDFs of $u_{syn}$, which belong to $\langle \alpha \rangle$= -15$^\circ$, 0$^\circ$, 30$^\circ$, 60$^\circ$ ($\sigma(\alpha)=12^\circ$) as well as the reference PDF $p_{ref}$. 
It becomes obvious that the synthetic PDFs match the reference PDF differently well.
To quantify the differences in these PDFs a weighted mean square error function in logarithmic space is considered, cf.  \cite{Feller1968},
\begin{equation}
\label{eq:KLE}
\epsilon(p_{syn},p_{ref}) =\frac{\int^{+\infty}_{-\infty}{(p_{syn}+p_{ref})(\ln(p_{syn})-\ln(p_{ref}))^2 \ du}} {\int^{+\infty}_{-\infty}{(p_{syn}+p_{ref})(\ln^2(p_{syn})+\ln^2(p_{ref})) \ du}}.
\end{equation}
$\epsilon$ is a natural logarithmic distance measure between $p_{syn}$ and $p_{ref}$. 
Figure \ref{fig:example2PDF} b) shows $\log(\epsilon)$ (logarithm of the basis 10) of $p_{syn}$ and $p_{ref}$ over $\langle \alpha \rangle$. A minimal distance at $\langle \alpha \rangle= 0^\circ$ becomes visible.
\\
Considering additionally dependency on the standard deviation of flaps motion, an ensemble of synthetic velocity time series is generated, which belongs to different combinations of $\langle \alpha \rangle$ and $\sigma$. 
To every single synthetic time series the PDF $p_{syn}(\langle \alpha \rangle,\sigma)$ is determined. 
Afterwards, these $p_{syn}$ and the reference PDF $p_{ref}$ are compared by means of $\epsilon$. $\log(\epsilon)$ as function of $\langle \alpha \rangle$ and $\sigma$ is shown in figure \ref{fig:example2epsilon}. 
In this figure \textit{every point} of the contour plot represents one calculated synthetic time series for a fixed set of excitation parameters $\langle\alpha\rangle$ and $\sigma$.
Isolines surround surfaces with \mbox{$\log(\epsilon)\le$ -3.5,   -1.7,   -0.87,   -0.43,   -0.21 and  -0.11.} 
Dark grey belongs to PDFs with a high agreement and bright grey to little similarity. 
The minimum of $\epsilon(\langle \alpha \rangle, \sigma)$ is located at the motion parameters $\langle\alpha\rangle=0^\circ\pm6^\circ$ and  $\sigma=12^\circ\pm3^\circ$. 
These parameters coincide with the motion parameters of the reference PDF $\langle\alpha\rangle=0^\circ$ and  $\sigma=11^\circ$, mentioned in section \ref{sec:Examples_1}.
Thus, we conclude from this example that the approach enables to find the proper magnitude of motion parameters to a given flow feature.
\\
We would like to give a few comments on this example.
First, figure \ref{fig:example2epsilon} presents an interesting aspect, namely that parameter sets can be found, which lead to similar flow features, e.g. in the trivial case $\langle\alpha\rangle=0^\circ\hat{=}180^\circ$ as well as how sensitive the flow feature is on the specific choice of the motion parameter.
Second, in addition to the estimation of motion parameters magnitude, one can consequently use this procedure to find relevant motion parameters and also suited motion patterns. 
Likewise to this example one can choose other parameters or motion patterns, 
which might modulate the flow in a suited manner. 
A map like figure \ref{fig:example2epsilon}, shows how strong chosen parameters and the pattern modulate the flow with regard to a considered flow feature.
However, one has to reflect on the issue of proper parameters.
For a reproduction of velocity PDFs the mentioned parameter, $\langle\alpha\rangle$ and $\sigma$ are proper parameters.
For a reproduction of power density spectra one might additionally consider a variation of the length of modulation, to get access to the frequency space.
Therefore, the parameter study has to be extended to more than two motion parameters.
Third, this parameter study by means of a computer is much faster than a conventional wind tunnel experiment.
This parameter study took less than 10 minutes (on an ordinary computer), which corresponds to a wind tunnel measurement period of roughly 50 hours! (15 different $\sigma$, 60 different $\langle \alpha \rangle$ and 200 s time series $\rightarrow$ 50 hours).

\subsection{Example: wind tunnel flow according to a reference flow}
 \label{sec:Examples_3}
\begin{figure}
\centering
\includegraphics[width=0.6\textwidth]{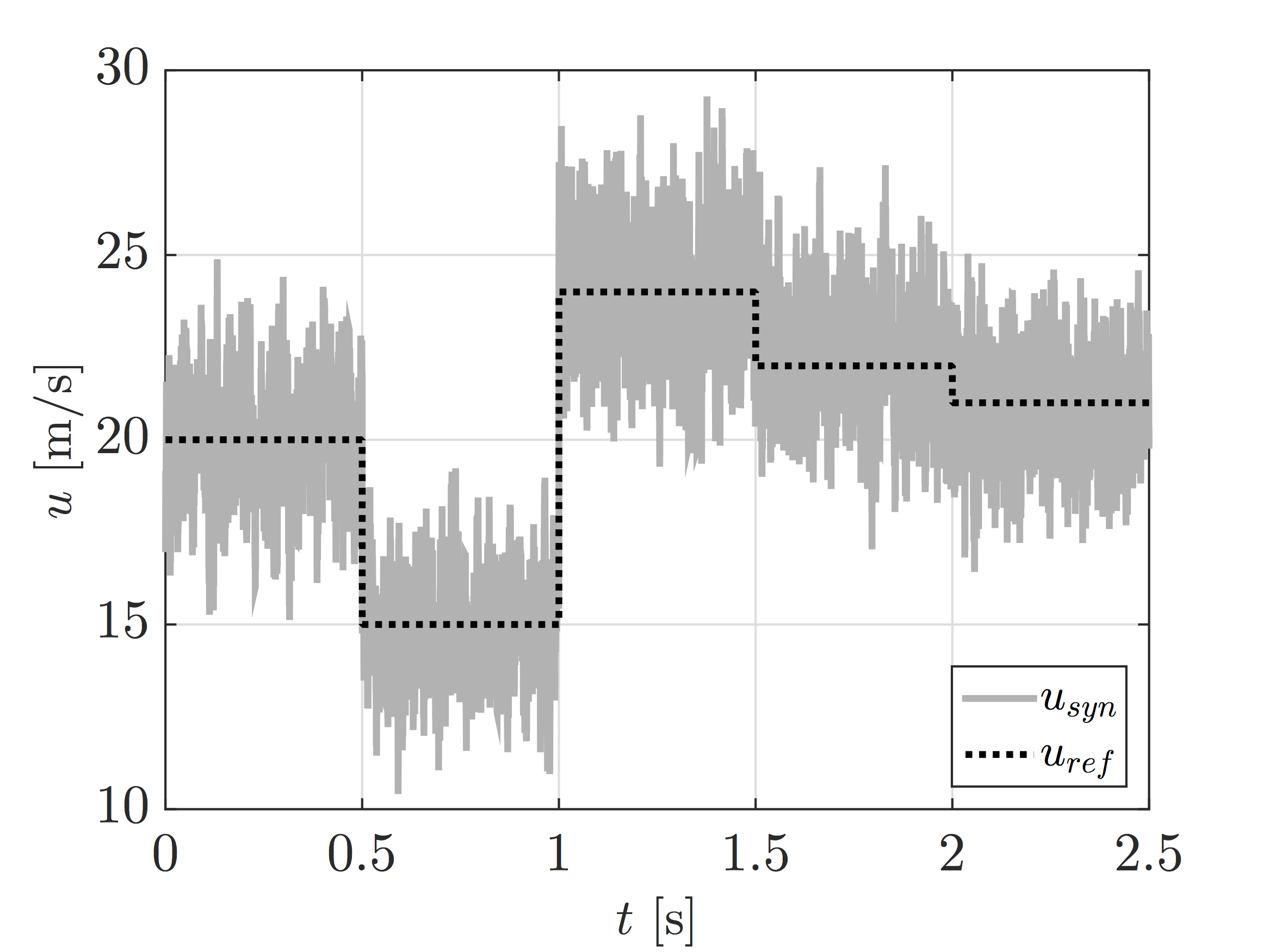}
\vspace{0.5cm} \caption{Synthetic velocity time series $u_{syn}$ according to a given reference velocity time series $u_{ref}$}
\label{fig:AG_Chara_von_u_ref_zu_u}
\end{figure}

\begin{figure}
\centering
\includegraphics[width=0.6\textwidth]{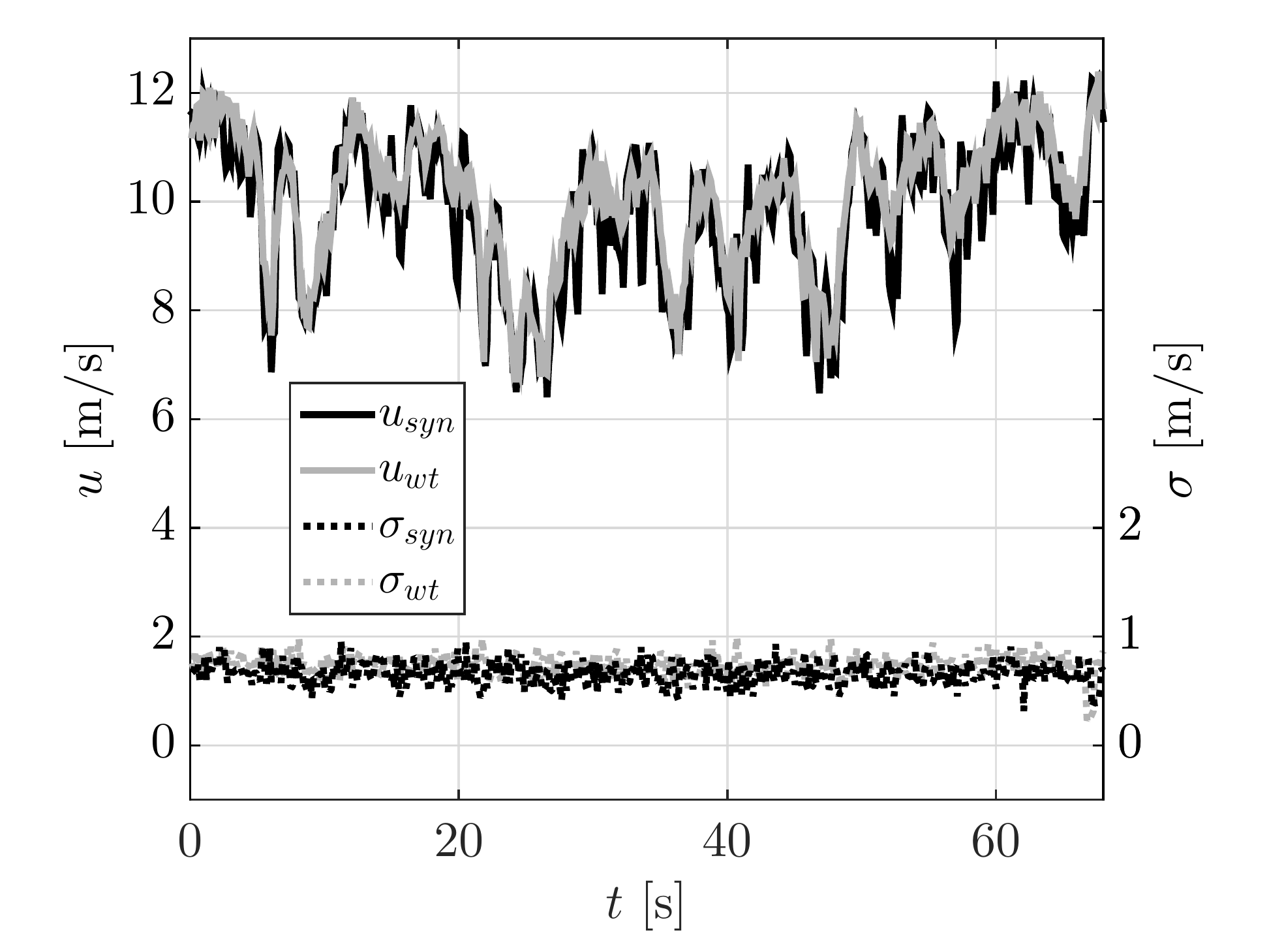}
\vspace{0.5cm} \caption{Synthetic (as well as reference) velocity time series and corresponding wind tunnel measured time series $u_{wt}$ as well as their moving standard deviations $\sigma_{syn}$ and $\sigma_{wt}$}
\label{fig:example_3}
\end{figure}

\begin{figure}
\centering
\includegraphics[width=0.6\textwidth]{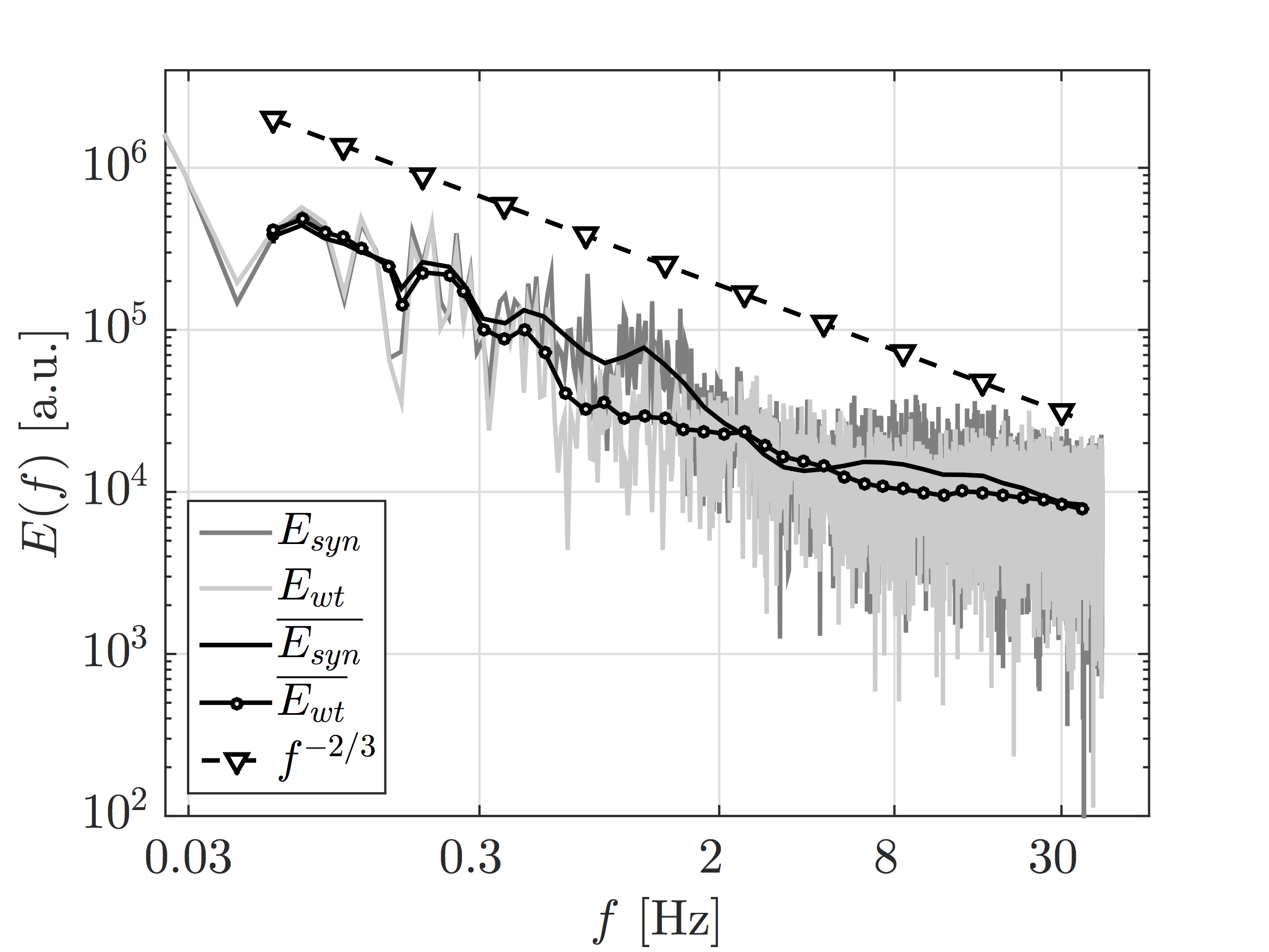}
\vspace{0.5cm} \caption{Power spectral density $E_{syn}$ and $E_{wt}$  of $u_{syn}$ and $u_{wt}$ shown in figure \ref{fig:example_3}. 
Additionally, window averaged spectra are shown, $\overline{E_{syn}}$ and  $\overline{E_{wt}}$.  $f^{-2/3}$ indicates the decrease of spectra}
\label{fig:example_3_2}
\end{figure}

This example shows the generation of synthetic velocity time series as well as the derivation of the flaps motion to a given reference velocity time series. 
To explain the procedure a simplified example is shown in figure \ref{fig:AG_Chara_von_u_ref_zu_u}. 
Here, the dotted line is an artificial velocity time series, which serves as reference time series $u_{ref}$. 
It consists of various velocities, which stay constant for $\Delta t = 0.5$~s. 
For the generation of a corresponding synthetic time series the following has to be done.
First, we consider the first 0.5~s of the reference velocity series, which is $u_{ref}=20~\frac{\rm{m}}{\rm{s}}$. 
From the wake calibration (fig. \ref{fig:AG_Chara_wake}) one finds the corresponding fluctuation, cut out a proper piece of fluctuations with a time length of $\Delta t = 0.5$~s. 
At the same time the corresponding $\alpha$ is specified, which determines the flaps positions which results in $u_{ref}=20~\frac{\rm{m}}{\rm{s}}$.
Adding to the piece of fluctuations the reference velocity $u_{ref}=20~\frac{\rm{m}}{\rm{s}}$, the first 0.5~s of the synthetic velocity time series $u_{syn}$ is generated in figure \ref{fig:AG_Chara_von_u_ref_zu_u}. Continuing this procedure for all reference velocities the complete synthetic velocity time series is generated by sticking all pieces in a row as shown in \ref{fig:AG_Chara_von_u_ref_zu_u}. Finally, $u_{syn}$ gets low passed filtered at $f_l$, see section \ref{sec:Examples_1}.
\\
Note that the procedure of generating such a synthetic velocity time series returns mainly two things. 
First, it shows to a certain degree how well the reproduction might be. 
For instance, figure \ref{fig:AG_Chara_von_u_ref_zu_u} shows that $u_{syn}$ is much more fluctuating than $u_{ref}$, thus differences in their stochastic feature are expected.
Second, by reproducing a reference signal also the flaps motions $\alpha(t)$ are specified. 
\\
Next, we apply this procedure on a \textit{real} reference velocity time series. 
The reference dataset is a free field measurement (by LiDAR - light detection and ranging) provided by \cite{Dooren2014} at DTU Ris{\o}.
Note that according to the maximal velocity in the reference data another wake calibration, different to figure \ref{fig:AG_Chara_wake}, is recorded with inflow velocity \mbox{$u_{\infty}=14~\frac{\rm{m}}{\rm{s}}$}.
The reference time series contains roughly 90 min of velocity measurement, sampled with approximately 100~Hz. 
For the reproduction, the reference data undergoes a window averaging, thereby reducing the number of velocity samples from \mbox{564 000} to 564.
The repetition time of the reference time series can be selected arbitrarily in the wind tunnel, it is not fixed to its real duration. 
Note that in some cases it can be beneficial to compress or to stretch the time series, e.g. comparison of different sized objects.
We reproduced the 90~min reference measurement in $\Delta T = 68$~s.
This has the consequence that the active grid modulates the flow with $f=\frac{564}{68\rm{s}}\approx 8.3$~Hz \mbox{($\sim t \approx 0.12$~s),} which is approximately four times faster than in the previous examples ($f=2$~Hz), section \ref{sec:Examples_1} and \ref{sec:Examples_2}.
\\
The synthetic velocity time series $u_{syn}$ and the measured time series $u_{wt}$ are shown in \mbox{figure \ref{fig:example_3}}. 
These time series have undergone a window averaging, thus the sample number is reduced to 564, which corresponds to the reduction of the reference velocity time series. 
$u_{wt}$ and $u_{syn}$ coincide on large time scales, 
but the quality of signal reproduction decreases to smaller time scales. 
Time series moving standard deviations of $\sigma_{syn}$ and $\sigma_{wt}$  is of a similar magnitude.
\\
To quantify deviations of $u_{wt}$ and $u_{syn}$ the power spectral density $E(f)$ over frequency $f$ of these signals is compared in 
figure \ref{fig:example_3_2}.
For the calculation of $E(f)$ the number of time series samples is only reduced to 5640, hence weaker averaged.
Thus, also higher frequencies can be compared.
Additionally, figure \ref{fig:example_3_2} shows window averaged power spectral densities $\overline{E}$, which facilitates the comparison. 
Overall, $\overline{E_{syn}}$ and $\overline{E_{wt}}$ show a good agreement.
In particular, they coincide for low frequencies up to  $f\approx0.3$~Hz. Thus, large flow modulations are correctly estimated.
The strongest deviations become apparent in the frequency range between \mbox{0.3~Hz $<f<$ 2~Hz,} which might be related to the decay of flow modulations, see section \ref{sec:Dynamic}. 
For the estimating approach one might conclude from this interpretation that \textit{shorter} flow modulations have to be proper \textit{amplified} on their initialization at the active grid, in such a way that these modulations have the desired magnitude at a certain downstream position.
Furthermore, different travel time of various flow modulations might be considered as well.
In contrast to example \ref{sec:Examples_1},
modulation length ($\sim t \approx 0.12$~s) is faster and position of measurement is further downstream. 
Thus, a different agreement in spectral features might be explained, see \ref{fig:example_1} a) vs. \ref{fig:example_3_2}.
Interestingly, the spectra converges again at highest frequencies.
An interpretation of this convergence might be that the amplitude of these flow modulations becomes small in comparison to the wake induced turbulence of the flaps. 
Thus, induced wake turbulence covers and overlies features of short modulations.
\\
Finally, we would like to conclude that the approach enables to find proper flaps motions to a given reference velocity time series. 
We evaluate the quality of the reproduction of the time series as well as its spectral characteristics as remarkably good in relation to the simplicity of the applied approach. 

\section{Conclusions}
\label{sec:Discussion}
An approach is presented which estimates wake features of an active grid. 
Based on an investigation of the grid wake, important and negligible features are worked out.
We found that flow modulations obey an exponential decay in the active grid wake. 
Their relaxation length depends on the length and intensity of modulations. 
Thus, the decay can be neglected under the conditions of long and strong modulation in relation to the downstream position.
Furthermore, the interaction of modulations is investigated. 
The results indicate no strong interaction of flow modulations. 
Consequently, we concluded that in a first approximation dynamic wake features can be neglected for an estimating approach, as long as mentioned conditions are fulfilled. 
Hence, the active grid wake is considered as quasi static and the flow modulation of the grid in time can be considered as a superposition of quasi static states.
These quasi static states get extracted from an experimental characterisation.
\\
The validity of this approach is shown by means of three examples, which test different aspects of flow modulations. 
We concluded from the achieved results that the approach is very well suited for the considered wake estimations. 
However, the third example also shows that above mentioned conditions have to be respected. 
Here, first indications of approach limits become visible, which might be the starting point for an extension of the approach. 
\\
All in all, we see new possibilities rising with this approach.
The approach shows already how a measurement from free field can be reproduced in the wind tunnel.
Such alternating flow conditions are of utmost importance
for many investigations, since only such flow conditions enable a realistic characterisation.

\section*{Acknowledgments}
We kindly acknowledge  the helpful discussions with  A. Fuchs, H. Hei{\ss}elmann, G. Kampers, L. Kr\"oger and S. Rockel
as well as for the LiDAR data provided by \cite{Dooren2014} in cooperation with M. v. Dooren.
This work is partially supported by the DFG-grant Pe 478/14-1 and Federal Ministry for Economic Affairs and Energy-grant 0325601D.

\section{Appendix}
\begin{figure}
\includegraphics[width=0.9\textwidth]{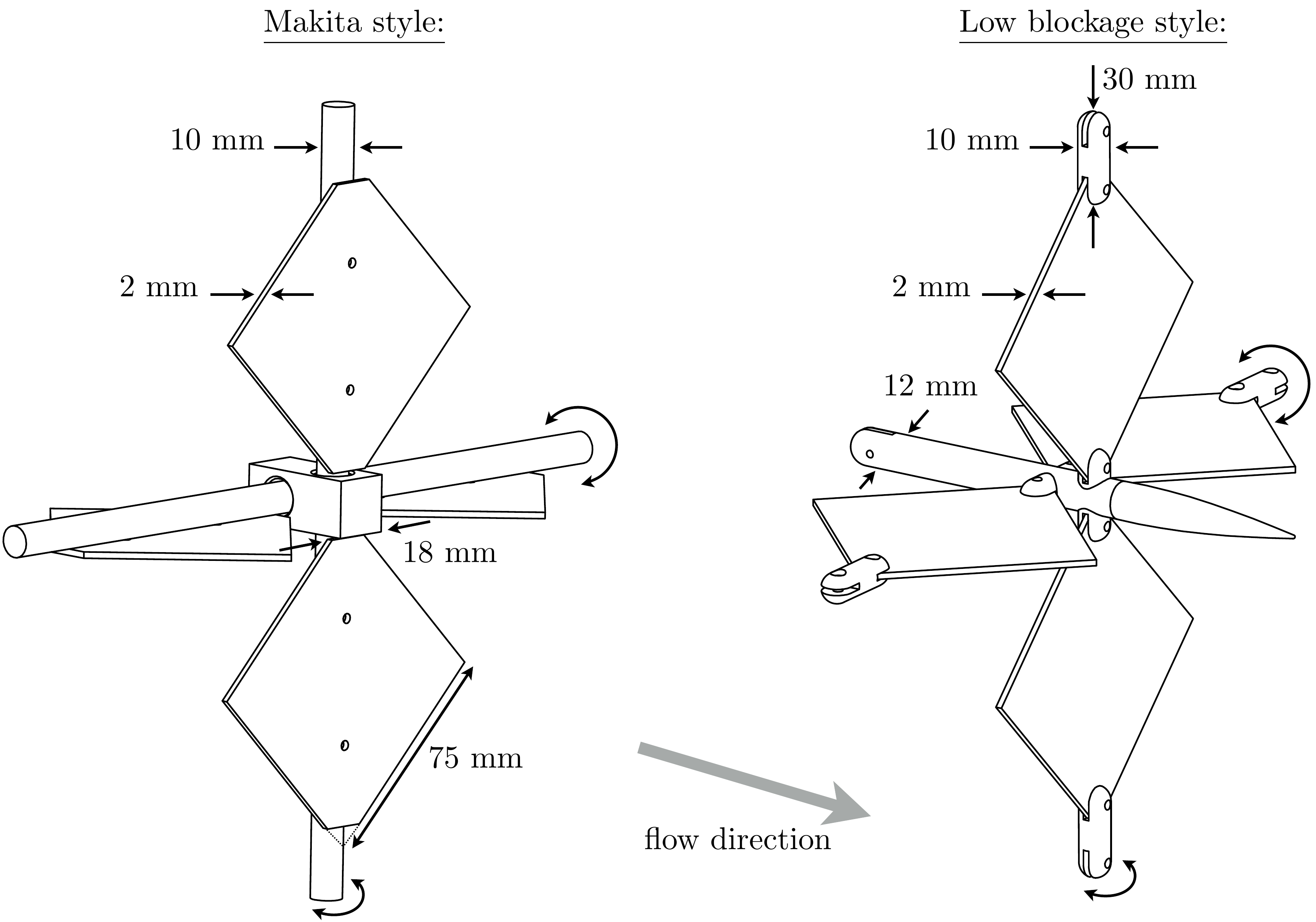}
\vspace{0.5cm} \caption{Detailed view of, standard active grid construction (left) and streamlined active grid design (right)}
\label{fig:design_vergleich}
\end{figure}

Figure \ref{fig:design_vergleich} shows two detailed views of the active grid. 
The left illustration presents the common active grid structure, where vertical and horizontal rods are support structures and the drive of flaps. Typically, vertical and horizontal rods are connected to achieve a higher grid stability.
This standard structure results usually in a minimal blockage of around 20\%, when the grid stands open for the incoming wind. 
In case the grid stands open, the flow is least disturbed, thus the highest flow velocity is present and lowest turbulence intensity, cf. \cite{Knebel2011}
 \mbox{$I(x=1.32$~m$)\ge5\%$}. 
 Consequently, higher flow speeds and lower turbulence intensities cannot be realised.
\\
To shift these limits, 
the idea to build a \textit{streamlined} and \textit{low blockage} active grid comes up.
The right part of figure \ref{fig:design_vergleich} shows a detail view of such a grid. 
The entire low blockage active grid is shown in figure \ref{fig:aG}.
Here continuous rods are avoided, only flaps with joints building up an axis.
Furthermore,  vertical and horizontal axes are connected by slim streamlined support structures, which minimally disturb the flow media. 
This concept leads to a minimal blockage of approximately 6\% and results in a minimal  turbulence intensity of \mbox{$I(x=1$~m$)\ge1.7\%$}, thus \textit{small} flow modulations become feasible. 
(This turbulence intensity is a mean of three measurements at representative wake positions, behind a support structure, behind a flap and between two flaps.)
\label{sec:Appendix}
\\
The minimal blockage can be determined as follows:
There are 
$7\times9$ 
support structures with a blocking area of 36$\pi$ mm$^2$, 
there are $2\times9\times7$ flaps with minimal blocking area of 2$\times$80 mm$^2$ 
and there are $2\times9\times7$ out looking adapters which connect the flaps with a blocking area of $200+25\pi - (36\pi-\frac{36\pi-100}{2})$~mm$^2$.
Thus, the total minimal grid blockage is roughly 47260~mm$^2$, which 
is equivalent to a blockage of  6.1\%, since the the nozzle outlet is $805\times1005$~mm$^2$.
The shown Makita style active grid has a minimal blockage of $\frac{7\times9\times18^2 + 2\times7\times9\times(92\times12)}{805\times1005}\approx19.7\%$. 
\\
We would like to comment two further positive aspects, which reveal in the ongoing work.
First, a low blockage active grid is especially beneficial when working with \textit{high} velocities. 
Since its pressure drop is low, we work with the grid up to a maximal speeds of $\sim$45$~\frac{\rm{m}}{\rm{s}}$, which comes close to the maximal wind tunnel flow velocity. 
Second, with a low blockage active grid also the angle of attack of flow can be modulated. 
However, first experiments indicate a high sensitivity on the rotational speed of flaps, therefore, the estimation of the angle of attack of flow is more delicate.


\bibliographystyle{plainnat}
\bibliography{P2arxiv_nur_pdfs}

\end{document}